\documentclass[reprint, prb, aps, floatfix, nofootinbib, superscriptaddress]{revtex4-2}
\usepackage{amsmath}
\usepackage{rotating}
\usepackage{graphicx}
\usepackage{chemformula}
\usepackage{mathtools}
\usepackage{tabularx}
\usepackage{color}
\usepackage{bm}
\usepackage{dcolumn}
\usepackage[colorlinks=true, linkcolor=blue, citecolor=blue, urlcolor=blue]{hyperref}

\renewcommand{\mathbf}[1]{\bm{#1}}

\begin{document}

\title{From continuum excitations to sharp magnons via transverse magnetic field\\ in the spin-1/2 Ising-like triangular lattice antiferromagnet \ch{Na2BaCo(PO4)2}}

\author{Leonie Woodland}
\affiliation{Clarendon Laboratory, University of Oxford Physics Department, Parks Road, Oxford OX1 3PU, UK}
\affiliation{ISIS Facility, Rutherford Appleton Laboratory, Chilton, Didcot OX11 0QX, UK}
\author{Ryutaro Okuma}
\affiliation{Clarendon Laboratory, University of Oxford Physics Department, Parks Road, Oxford OX1 3PU, UK}
\affiliation{Institute for Solid State Physics, University of Tokyo, Kashiwa, Chiba 277-8581, Japan}
\author{J. Ross Stewart}
\affiliation{ISIS Facility, Rutherford Appleton Laboratory, Chilton, Didcot OX11 0QX, UK}
\author{Christian Balz}
\thanks{Present address: Neutron Scattering Division, Oak Ridge National Laboratory, Oak Ridge, Tennessee 37831, USA}
\affiliation{ISIS Facility, Rutherford Appleton Laboratory, Chilton, Didcot OX11 0QX, UK}
\author{Radu Coldea}
\affiliation{Clarendon Laboratory, University of Oxford Physics Department, Parks Road, Oxford OX1 3PU, UK}

\begin{abstract}
We report high-resolution inelastic neutron scattering measurements of the excitation spectrum in large single crystals of the spin-1/2 triangular lattice Ising-like antiferromagnet \ch{Na2BaCo(PO4)2} in magnetic fields applied transverse to the Ising axis. In the high-field polarized phase above a critical field $B_{\rm{C}}$, we observe sharp magnons, as expected in the case of no exchange disorder. Through simultaneous fits to the dispersions including data in polarizing field along the Ising axis, we obtain an excellent match to an Ising-like XXZ Hamiltonian and rule out previously proposed Kitaev exchanges. In the intermediate-field phase below $B_{\rm{C}}$, we observe three dispersive modes, out of which only the lowest energy one is sharp and the others are broad and overlap with continuum scattering. We propose that the broadening effects are due to magnon decays into two-magnon excitations and confirm that such processes are kinematically allowed. The continuum scattering becomes progressively stronger upon lowering field and, at 0.25~T and zero field, it dominates the complete spectrum with no clear evidence for even broadened magnon modes. We discuss the relevance of the continuous manifold of mean-field degenerate ground states of the refined Hamiltonian for capturing the observed spectrum in zero field, and compare the data with the one- and two-magnon spectrum averaged over this manifold. We also propose a model of the interlayer couplings to explain the observed finite interlayer magnetic propagation vector of the zero-field magnetic order; this requires the breaking of the mirror symmetry in the nominal $P\bar{3}m1$ space group and through refinement of x-ray diffraction data on an untwinned single crystal, we indeed confirm a rotation of the CoO$_6$ octahedra around the $c$-axis, which lowers the symmetry to $P\bar{3}$.  
\end{abstract}

\date{September 12, 2025}
\maketitle
\section{Introduction}

The spin-1/2 triangular lattice antiferromagnet (TLAF) is a canonical example of a frustrated quantum spin model in two dimensions. A much explored theoretical scenario is the XXZ spin Hamiltonian   
\begin{equation}\label{E:XXZnbcpo}
\mathcal{H}_{\rm{XXZ}}=\sum_{\langle ij \rangle} J_{xy}\left(S^x_iS^x_j + S^y_iS^y_j\right) + J_z S^z_iS^z_j,
\end{equation}
with $J_z,J_{xy} > 0$ and the sum extending over all $ij$ nearest-neighbor bonds counted once.
For isotropic and easy-plane exchange ($J_z \leq J_{xy}$) the ground state has coplanar 120$^{\circ}$ order \cite{Singh1992,Bernu1994,Capriotti1999}, although with a much reduced ordered moment due to strong zero-point quantum fluctuations, whereas in the pure Ising limit ($J_{xy}=0$) a classical spin liquid is expected \cite{Wannier1950}. For the intermediate case of an Ising-like exchange ($J_z>J_{xy}$), the ground state and quantum excitation spectrum are still very much the subject of active research \cite{Fazekas1974,Kleine1992,Ulaga2024,gallegos2025phasediagrameasyaxistriangularlattice}, with work still ongoing to understand the spectrum and phase diagram in applied field \cite{Yamamoto2019,xu2024simulatingspindynamicssupersolid}, due in part to a lack of suitable experimental realizations until very recently.

\begin{figure}
\centering
\includegraphics[width=0.48\textwidth]{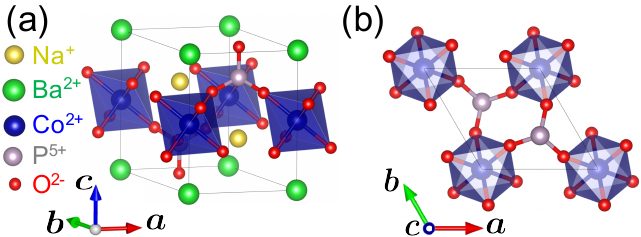}
\caption[The crystal structure of \ch{Na2BaCo(PO4)2}]{(a) Hexagonal crystal structure of \ch{Na2BaCo(PO4)2} as per Table~\ref{T:XRD_100K_P-3} in Appendix~\ref{A:xrays}, with \ch{Co^{2+}} ions inside O$_6$ octahedra (blue shading) arranged in triangular layers stacked along $c$. (b) Bonding of the octahedra in the $ab$ plane. The solid outline in both panels is the hexagonal unit cell of the $P\bar{3}$ space group.
}\label{F:structure}
\end{figure}

The material \ch{Na2BaCo(PO4)2} with \ch{Co^{2+}} moments arranged in vertically stacked triangular layers (see Fig.~\ref{F:structure}) \cite{Zhong2019} has been proposed to be a good realization of an effective spin-1/2 Ising-like XXZ TLAF \cite{Li2020aa,Gao2022aa,Sheng2022,mou2024comparative,Xiang2024aa}, as in Eq.~(\ref{E:XXZnbcpo}) 
with $z\parallel c$-axis. Crystallographically, the local $\bar{3}$ point group at each \ch{Co^{2+}} site ensures that all nearest-neighbor bonds are symmetry equivalent and the absence of structural disorder/site mixing effects makes this an ideal candidate to explore the clean quantum limit. 
Furthermore, the energy scale of the interactions is small enough that the transitions to the polarized phases for field both along and transverse to the Ising axis occur at sufficiently low magnetic fields that inelastic neutron scattering (INS) measurements of the dispersion relations can be performed \cite{Li2020aa}; by analyzing such data one can quantitatively deduce the spin Hamiltonian \cite{Coldea2002}. 
All these features make this material an ideal candidate for the experimental investigation of the whole phase diagram of the quantum Ising-like XXZ TLAF. In comparison to another recently proposed realization of this model, \ch{K2Co(SeO3)2} \cite{zhu2024continuumexcitationsspinsupersolidtriangular,chen2024phasediagramspectroscopicevidence}, the current material has a smaller, but still sizeable Ising anisotropy  ($J_z/J_{xy}=1.6$ compared to 12) and a smaller exchange energy scale. 

\begin{figure}[tb]
\centering
\includegraphics[width=0.5\textwidth]{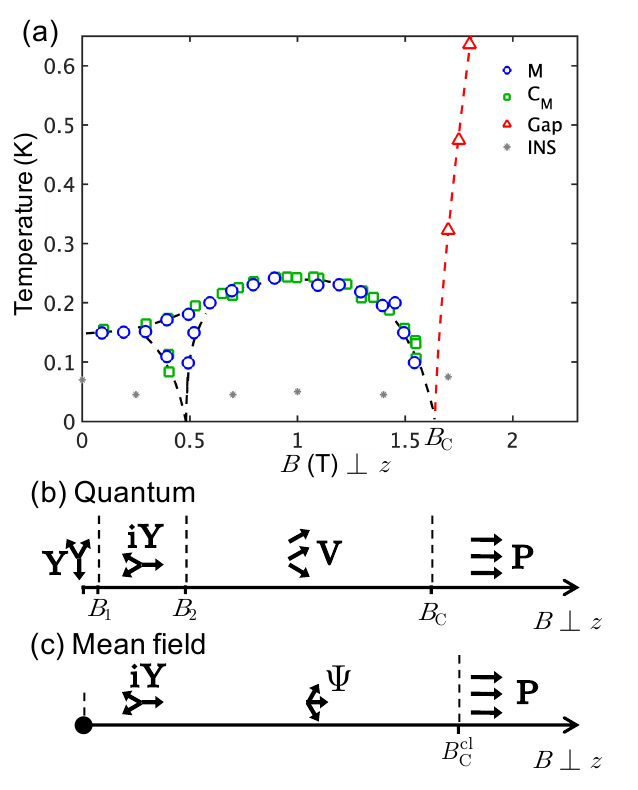}
\caption[Phase diagram of \ch{Na2BaCo(PO4)2} in in-plane field]{(a) Experimental phase diagram of \ch{Na2BaCo(PO4)2} for $B\!\perp \!z$. Dashed black lines are guides to the eye for phase boundary lines defined by anomalies in heat capacity (green squares) and magnetization (blue open circles) from Ref.~\cite{Sheng2022}. 
Red triangles show the estimated magnon gap, expected to close at the critical transition field $B_{\mathrm{C}}$ (dashed red line is a guide to the eye); the gap was estimated from the observed minimum gap to excitations in single orientation INS data. Gray stars indicate where the INS data in Fig.~\ref{F:twomagnon} were collected. [(b), (c)] Theoretically expected quantum/mean-field phase diagrams, respectively, for $B\perp z$ \cite{Yamamoto2019,Gao2022aa}, for the XXZ model in Eq.~(\ref{E:XXZnbcpo}). The solid dot $\bullet$ symbol in (c) at $B=0$ indicates a ground state degeneracy at the mean field level between many 3-sublattice structures, which is lifted by an infinitesimal applied field. Black arrows show the orientation of the spins of the three magnetic sublattices, each occupying one of the three corners of every triangle in the $ab$ plane as described in the text; the convention used here and in Fig.~\ref{F:twomagnon} is that the Ising direction is vertical while the field points to the right. The fields in panels (b) and (c) are not drawn to scale.}\label{F:phasediagram}
\end{figure}

While previous works on \ch{Na2BaCo(PO4)2} focused on thermodynamic measurements across the phase diagram \cite{Zhong2019,Li2020aa,huang2022thermal,Liu2022aa,xu2025nmrstudysupersolidphases} or on the spectrum in field along the Ising axis \cite{Sheng2022,Sheng2024,Popescu2025}, here we focus on the evolution of the spectrum in field applied transverse to the Ising axis. Above a critical field $B_{\rm C} \simeq 1.65$~T estimated from thermodynamic measurements \cite{Lee2021,Sheng2022}, we observe sharp magnons, with dispersions well described by the XXZ Hamiltonian in Eq.~(\ref{E:XXZnbcpo}). Through simultaneous fits to the dispersion relations observed at multiple field values, as well as to previously reported dispersions in the polarized phase for field along the Ising axis \cite{Sheng2022}, we refine all parameters of the XXZ Hamiltonian and $g$-tensor, and place an upper bound on additional couplings beyond this minimal model. This quantitative characterization of the Hamiltonian is crucial for comparisons of the low-field excitation spectrum to theoretical models. 

Below the critical field $B_{\rm C}$, an inelastic scattering continuum becomes clearly apparent, and its weight increases strongly upon decreasing field. Two distinct magnetic phases are experimentally observed at the lowest temperatures, see Fig.~\ref{F:phasediagram}(a). In the phase immediately below $B_{\rm C}$, the spectrum evolves smoothly upon decreasing field from the spectrum above $B_{\rm C}$ via Brillouin zone folding. The magnons become increasingly broadened as the field is decreased, but the lowest energy magnons remain sharp, as expected in a scenario of magnon decays at intermediate and high energies. In contrast, in the low-field phase (probed at 0 and 0.25~T) no sharp modes are resolved at any of the probed energies, and the spectrum is instead dominated by a very broad excitation continuum, consistent with previously reported zero-field INS measurements \cite{Sheng2024,gao2024spinsupersolid}. The absence of well-defined magnons in this regime is very surprising given that the system shows long-range magnetic order manifested in well-defined magnetic Bragg peaks observed on the elastic line during the same measurement.    

The above behavior is also in stark contrast to the physics of the spin-1/2 TLAF with a weak easy-plane XXZ exchange anisotropy, as realized in \ch{Ba3CoSb2O9}, where sharp magnons are observed in zero field throughout the Brillouin zone with no magnon decays anywhere \cite{Macdougal2020}. In that case, the absence of magnon decays was attributed to the presence of strong quantum interactions between magnons and the higher-energy continuum excitations \cite{Verresen2019le}, which renormalize the magnon energies downwards to keep them below the lower boundary of the excitation continuum, such that decay effects become kinematically disallowed. Completely different physics seems to apply in the case of Ising-like exchange anisotropy relevant for the present material, in which magnon decays become apparent immediately upon lowering the field below $B_{\rm C}$ and continuum scattering increases strongly in intensity upon lowering field, dominating the full spectrum at zero and low field. 

The remainder of this paper is organized as follows. Section~\ref{S:expdetails} outlines the method used to grow large single crystals and the details of the INS experiments to probe the excitations. The following Sec.~\ref{S:highINS} reports INS results in the high-field polarized phase and the global fits used for the quantitative determination of the spin Hamiltonian. Sec.~\ref{S:phasediagram} reviews the theoretically expected phase diagram in transverse field in Fig.~\ref{F:phasediagram}(b)-(c), then Sec.~\ref{S:lowINSexpcalc} reports the INS spectrum as a function of transverse field, which is compared quantitatively with predictions of linear spinwave theory (LSWT) for the refined Hamiltonian, including both one- and two-magnon excitations. Sec.~\ref{S:degeneracy} discusses the possible relevance of the continuous manifold of mean-field degenerate ground states of the refined Hamiltonian for capturing the spectrum in zero field and compares the data with the expected one- and two-magnon spectrum averaged over such a manifold. Sec.~\ref{S:interlayer} presents the magnetic diffraction pattern in zero field and proposes a model of interlayer couplings to explain the observed finite magnetic propagation vector along the interlayer direction. For completeness, Sec.~\ref{S:diffractionB} presents the evolution of the magnetic diffraction pattern in transverse field. Finally, Sec.~\ref{S:conclusion} summarizes the main results and conclusions. 

The Appendices present crystal structure refinements and technical details of the spinwave calculations supporting the analysis. Appendix~\ref{A:xrays} reports x-ray diffraction on both twinned and untwinned single crystals and proposes a revised crystal structure that breaks the vertical mirror planes, as required by the model of interlayer couplings discussed in Sec.~\ref{S:interlayer}. Appendix~\ref{A:JKfits} presents spinwave dispersions for an alternative Heisenberg-Kitaev model and discusses why this is ruled out by the present high-field INS data. Appendix~\ref{A:2M} outlines the method used in the quantitative calculation of the INS two-magnon scattering intensity for the 3-sublattice non-collinear orders expected at zero and intermediate transverse field. Appendix~\ref{A:Vphase} compares the spectrum for two distinct ground state models below $B_{\rm C}$, the mean-field $\Psi$-phase and the quantum-fluctuation stabilized V-phase. Appendix~\ref{A:NBCPO_interlayer} discusses the spinwave spectrum in the polarized phase above $B_{\rm C}$ for a model that includes the (weak) interlayer couplings and proposes an estimated upper bound of their magnitude.

\section{Experimental details}\label{S:expdetails}

Centimeter-size single crystals were grown using a flux method modified from that reported in Ref.~\cite{Zhong2019} using a seed crystal. Na$_2$CO$_3$, BaCO$_3$, Co$_3$O$_4$, (NH$_4$)$_2$HPO$_4$, and NH$_4$Cl were mixed in 1.1:1:1/3:2:1.2 molar ratio with a typical total mass of 50~g and thoroughly ground by ball milling. The ground powder was loaded in an alumina crucible and calcined at 900$^\circ$C for 24 hours to obtain Na$_2$BaCo(PO$_4$)$_2$ powder. The powder was mixed with flux media, NaCl and MoO$_3$, in a molar ratio of 1:3:0.1. The mixed powder was loaded into a Pt crucible with 30~ml capacity. The crucible was fully filled and heated to 900$^\circ$C, kept at this temperature for 2~hours, then cooled to room temperature and refilled with mixed powder, and the heating and refilling cycle repeated until the top melt surface left about 1~cm from the top for a mm-size seed single crystal of Na$_2$BaCo(PO$_4$)$_2$ to be placed in this space inside the crucible but above the melt. The crucible was covered with a Pt foil with a hole in the center to insert the seed crystal. The seed crystal was suspended by a Pt wire from the chimney of the box furnace and initially placed below the Pt foil, but above the melt. The Pt crucible and the reaction mixture were then again heated to 900$^\circ$C, kept at this temperature for 2~hours, and cooled to 850$^\circ$C. At that point, the seed crystal was lowered into the melt by manually changing the height of the Pt wire, and then the temperature was slowly reduced to 800$^\circ$C at a rate of  -2$^\circ$C/hour, after which the furnace was cooled to room temperature by switching off the heater. The temperature ramping rate was $\pm$100$^\circ$C/hour unless stated otherwise. After the reaction, the mixture was washed with water. Centimeter-size, dark pink crystals, with a hexagonal plate morphology in the $ab$ plane, grew mainly near the surface, but also from the wall of the crucible. The orientation of each crystal was checked by single crystal X-ray diffraction by sampling several points on the edges; for more details see Appendix~\ref{A:xrays}.

\begin{figure}[htb]
\centering
\includegraphics[width=0.48\textwidth]{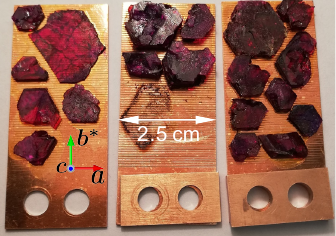}
\caption{Picture of the co-aligned single crystals of \ch{Na2BaCo(PO4)2} glued on oxygen-free copper sheets, before being assembled in the sample mount used in the INS experiments. 
}\label{F:samplemount}
\end{figure}

Inelastic neutron scattering (INS) measurements of the magnetic excitation spectrum were performed on a sample of 19 co-aligned single crystals totaling 6.41~g, glued on oxygen-free copper sheets as illustrated in Fig.~\ref{F:samplemount}. The crystals were mounted with the $c$-axis normal to the copper sheets and the $a$-axis horizontal, as indicated in the figure, with the measured overall mosaic $3^{\circ}$ (full width half maximum, FWHM). The INS measurements were performed using the direct geometry time-of-flight spectrometer LET \cite{LET} at the ISIS facility. The sample assembly was mounted upside-down in relation to the photograph in Fig.~\ref{F:samplemount} inside a vertical cryomagnet and magnetic fields up to 7~T were applied along the $b^*$ direction, i.e. within the plane perpendicular to the Ising axis, and such that the horizontal scattering plane was the crystallographic $ac$ plane [reciprocal $(h,-0.5h,l)$ plane]. Throughout this paper, we express wave vectors as $(h,k,l)$ in terms of reciprocal lattice units (r.l.u.) of the structural hexagonal unit cell. The magnet was placed in the instrument such that the support wedges allowed unimpeded travel for incident neutrons to scatter horizontally in the range $55^{\circ}<2\theta<135^{\circ}$, such that it was possible to access relatively large wave-vector transfers using small incident neutron energies and consequently achieve a high energy resolution. The sample was cooled using a dilution refrigerator insert and all data were collected below 80~mK, well below the zero-field ordering transition temperature of $\approx 148$~mK \cite{Li2020aa}, except some measurements in the high-field polarized phase at 3.5~T, collected below 165~mK, still cold enough to be in the low-temperature limit at this field where the magnon gap is $\approx 0.46$~meV $\approx 5.3$~K. LET was operated to measure simultaneously the inelastic scattering of incident neutrons with energies of $E_i = $ 0.87, 1.41 and 2.67~meV; the measured energy resolutions (FWHM) on the elastic line were 0.032(2), 0.060(3) and 0.143(2)~meV, 
respectively. The elastic line was centered at zero energy transfer to within better than 1 $\mu$eV at every energy probed.

Multi-angle (Horace) scans were collected in order to obtain full four-dimensional data sets of the scattering intensity as a function of energy and momentum transfer. In a first experiment, measurements were taken at 0, 1.7, 3.5 and 7~T, while in a second experiment, measurements were taken at 0.25, 0.7, 1, 1.4 and 3.5~T, with the repeated 3.5~T measurement used for comparison between the two data sets and for background subtraction. We estimate that the overall scale of the magnetic scattering intensity was consistent between the two experiments to within 10\%. 
For the above measurements, the sample was rotated about the $b^*$ axis through an angular range of $180^{\circ}$ in steps of $2^{\circ}$. Each orientation was counted for a typical counting time of 7 $\mu$Ah of protons on target (at an average proton current of 40 $\mu$A). 
Additional measurements to observe the dependence of the gap on field were collected at finely spaced fields between 1.55 and 3.5~T for a fixed sample orientation [with the (1,-0.5,0) axis at $3^{\circ}$ relative to the incident beam direction] with a typical counting time of 18~$\mu$Ah.  
 The raw time-of-flight neutron data were converted to scattering intensities $S(\mathbf{Q},\omega)$ using \textsc{mantid} \cite{Arnold2014}. 
 The data were then analyzed using the \textsc{horace} \cite{Ewings2016} and \textsc{mslice} \cite{MSlice} packages.

To obtain the magnetic inelastic intensities, an estimate of the non-magnetic background signal was subtracted off. This was constructed from the observed scattering intensities at wavevectors and energy transfers where no inelastic magnetic scattering is expected, such as below the single-magnon dispersion in the polarized phase at high field, or from the data at intermediate fields at wavevectors and low energies below the dispersion of the lowest energy sharp mode, in cases when this mode could be clearly resolved from the elastic line.

\section{Excitations in the polarized phase above $B_{\rm{C}}$}\label{S:highINS}
\begin{figure}[tb]
\centering
\includegraphics[width=0.5\textwidth]{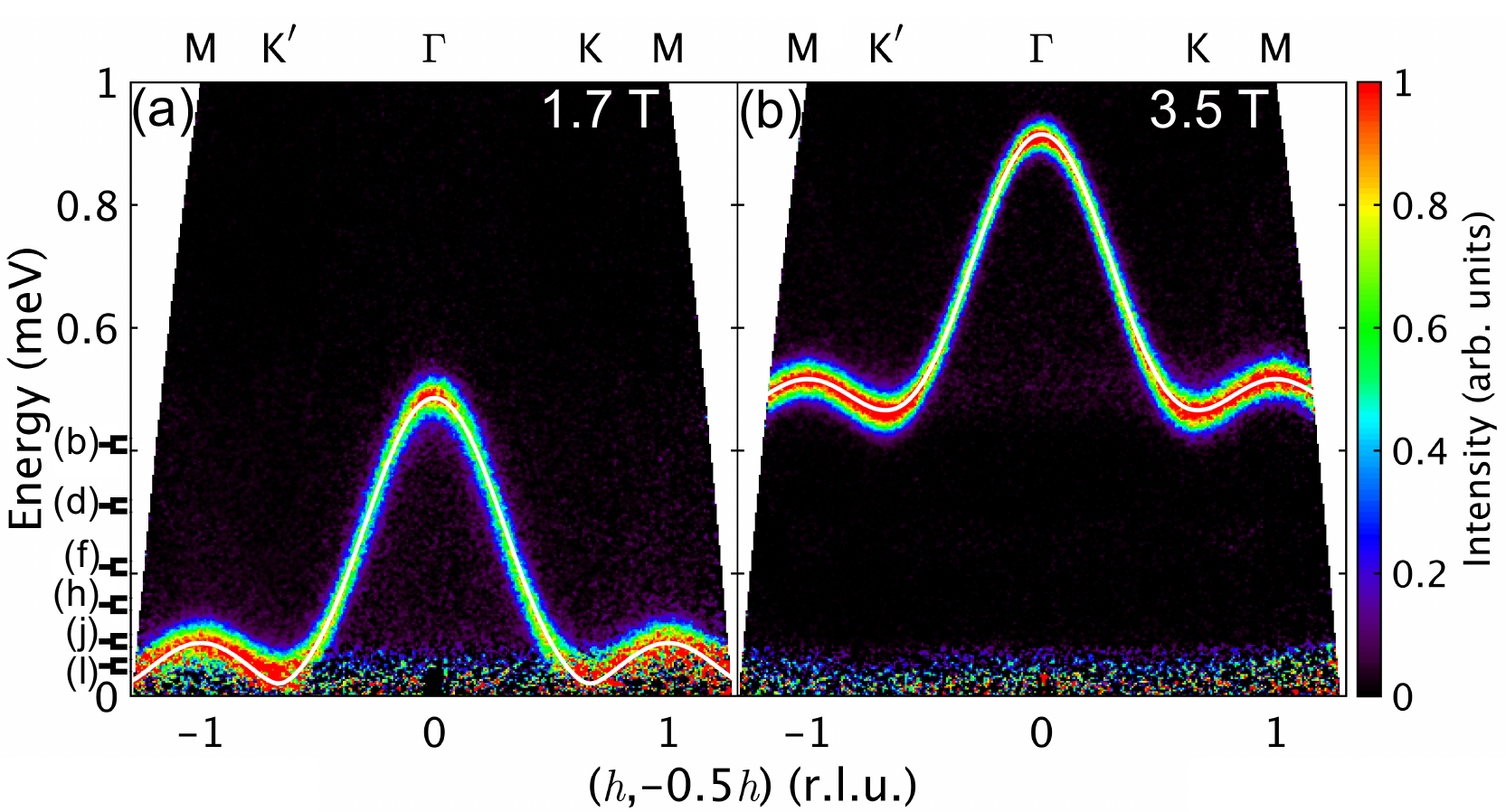}
\caption[Spectrum of \ch{Na2BaCo(PO4)2} in transverse field in the field polarized phase]{
INS spectrum along the in-plane (1,-0.5) direction in the field-polarized phase just above $B_{\rm{C}}$~(a) and at 3.5~T~(b). Intensities were averaged in the range $|k|<0.025$ in the $(0k0)$ direction and across the full range of data along $(00l)$, approximately $-1.7<l<0.7$. A sharp dispersive mode is seen in both cases, as expected for coherently-propagating magnons in the field-polarized phase. The white curves are the calculated magnon dispersion in Eq.~(\ref{E:lswtB}) using the best-fit parameters in Table~\ref{T:hamparams}. The brackets labeled (b) to (l) on the left-hand side of (a) indicate the averaging range in energy for the corresponding $hk$ slices in Fig.~\ref{F:constantenergy}. In both panels, color indicates the INS intensity on an arbitrary scale, 
after subtracting off the 7~T data, which was considered to be a good estimate of the non-magnetic background signal as at this high field no magnetic inelastic signal is expected over the whole plotted energy range since the magnon gap exceeds 1.3~meV. The incident neutron energy was $E_i=1.41$~meV. Labels on the top horizontal axis indicate positions equivalent to high symmetry points in the two-dimensional hexagonal Brillouin zone as indicated in Fig.~\ref{F:constantenergy}(b). 
}\label{F:highfieldresults}
\end{figure}

In the polarized phase above $B_{\rm C}$, we observe sharp magnon excitations as illustrated in Fig.~\ref{F:highfieldresults}. The magnon linewidth is essentially resolution-limited, with no evidence for any intrinsic width.  
This is in agreement with the expectation that there is no structural disorder in the present sample.

To parameterize the empirical dispersion relations, Gaussian peak shapes were fit to energy scans through the Horace INS data at several fields (1.7, 3.5 and 7~T) to extract several hundred dispersion points. For the comparison with the expected magnon dispersions for model spin-exchange Hamiltonians, the Zeeman interaction was parameterized as 
\begin{equation}
\mathcal{H}_{\rm{Zeeman}}=- \mu_B\sum_i \left(g_{ab}B_{y}S^y_i+g_cB_zS^z_i\right),  \label{E:Zeeman}    
\end{equation}
for an applied magnetic field with components along the $y$ and $z$ directions, with corresponding $g$-tensor components $g_{ab}$ and $g_c$, and where $x, y, z$ are along the $(1,-0.5,0)$, (010) and (001) directions, respectively. As discussed later in Appendix~\ref{A:NBCPO_interlayer}, no dispersion could be resolved along the interlayer $(001)$ direction, indicating that the 3D interlayer couplings are very small in magnitude, estimated to be at most a few percent of the exchange strength between nearest-neighbors in the triangular layers. Therefore, in the analysis of the inelastic spectrum in this and the following section, we consider a strictly two-dimensional spin-exchange model and plot the inelastic data in terms of a two-dimensional wave vector expressed as $(h,k)$ in r.l.u. components.          

\subsection{Fits to an XXZ Hamiltonian}\label{S:XXZfit}
We first compare the measured dispersions to predictions for the XXZ model in Eq.~(\ref{E:XXZnbcpo}), with an alternative model considered in Appendix \ref{A:JKfits}. The magnon dispersion for the full Hamiltonian $\mathcal{H}_{\rm{XXZ}}+\mathcal{H}_{\rm{Zeeman}}$ in the polarized phase in field along $z$ is 
\begin{equation}
\hbar\omega_{\parallel}(\mathbf{Q}) = 2S[g_c \mu_B B_z -3J_z + J_{xy}\gamma(\mathbf{Q})],\label{E:lswtC}
\end{equation}
where
\begin{equation}
\gamma(\mathbf{Q})=\cos{2\pi h}+\cos{2\pi k}+\cos{2\pi (h+k)}, \label{E:gamma} 
\end{equation}
with the effective spin $S=1/2$ for the Co$^{2+}$ ions in \ch{Na2BaCo(PO4)2}.
Note that only the $J_{xy}$ term contributes to the magnon bandwidth, whereas both the Zeeman term, whose strength is determined by the $g$-factor, and the Ising exchange $J_z$ have the effect of a uniform shift of the magnon energies. Therefore, to decouple the effects of the $g$-factor and of the Ising exchange $J_z$, more constraints are needed, such as dispersion measurements for at least two distinct fields in the polarized phase, or simultaneously including in the fit the dispersions in the polarized phase in transverse field, where the dispersion has a different functional dependence on the exchanges. In this latter case, the magnon dispersion is 
\begin{align}
\hbar\omega_{\perp}(\mathbf{Q})=&2S\sqrt{\mathcal{A}^2-\mathcal{B}^2},\label{E:lswtB}\\
\mathcal{A}=&g_{ab}\mu_B B_{y} - 3J_{xy} + \frac{J_z+J_{xy}}{2}\gamma(\mathbf{Q}),\nonumber \\
\mathcal{B}=&\frac{J_z-J_{xy}}{2}\gamma(\mathbf{Q}). \nonumber
\end{align}
Note that, in this case, both exchange terms $J_z$ and $J_{xy}$ now contribute to the magnon bandwidth. 

\begin{figure}[tbh]
\centering
\includegraphics[width=0.45\textwidth]{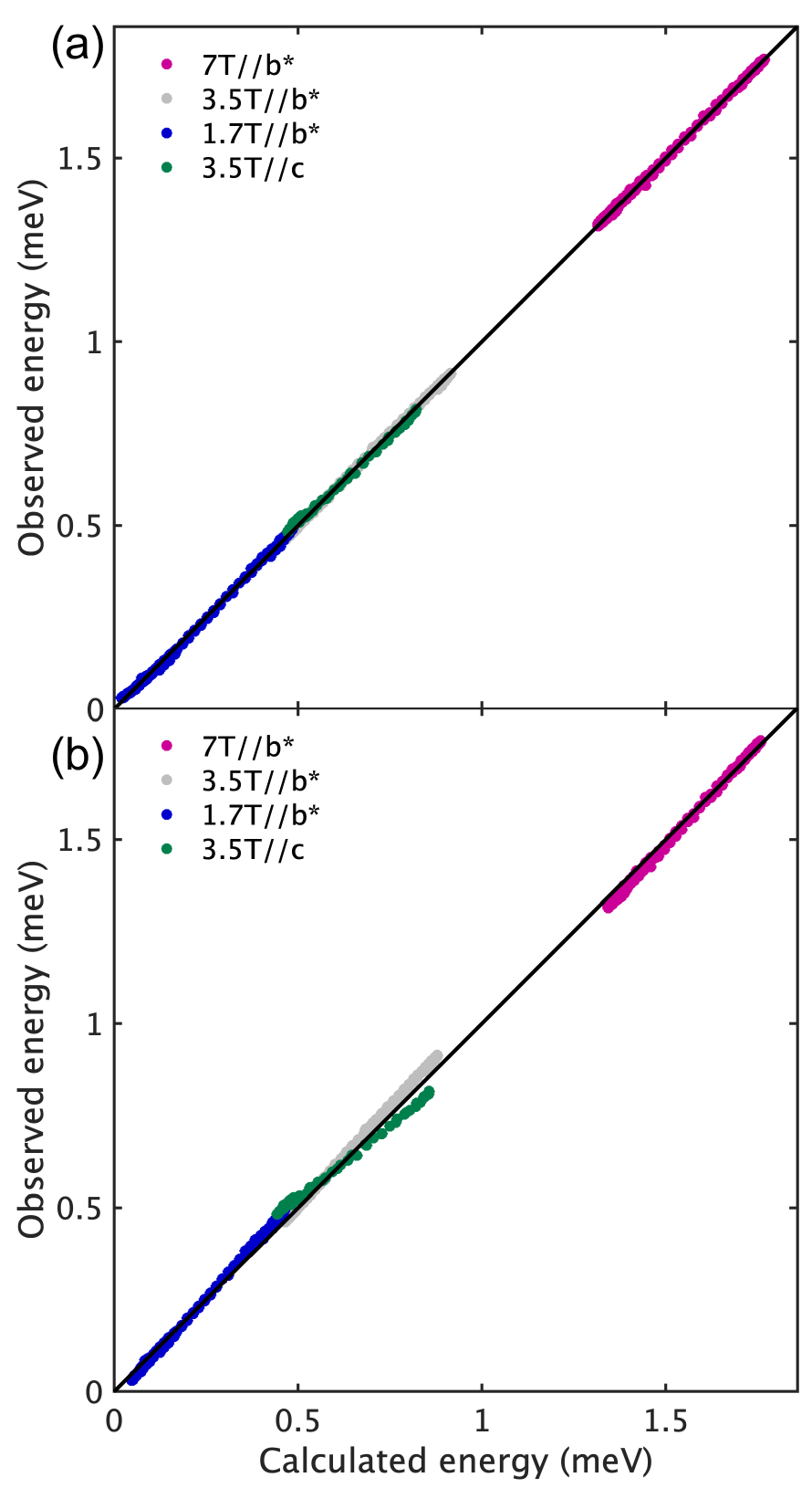}
\caption[Observed vs. calculated magnon energies for two Hamiltonian models for \ch{Na2BaCo(PO4)2}]{Observed vs. calculated magnon energies using different spin Hamiltonians. (a) The XXZ model in Eq.~(\ref{E:XXZnbcpo}) with best fit parameters in Table~\ref{T:hamparams} gives an excellent quantitative description (the solid line indicates the 1:1 agreement). (b) The Heisenberg-Kitaev model in Eq. (\ref{E:hamJK}) discussed in Appendix~\ref{A:JKfits}, with best fit parameters in Table~\ref{T:hamparamsJK}, shows clear disagreements. Different color symbols correspond to dispersion points collected at different field values or directions as indicated in the legend. Dispersion points for 3.5~T~$\parallel c$ were extracted by digitizing the data in Fig.~4B of Ref.~\cite{Sheng2022}. For all dispersion points, the uncertainty in the observed energy is smaller than the size of the symbols.}\label{F:onetoone}
\end{figure}

Fits were performed simultaneously to the empirically-extracted dispersion points at 3.5 and 7~T~$\parallel b^*$ and 3.5~T~$\parallel c$, the latter extracted by digitizing the data reported in Ref.~\cite{Sheng2022}, with excellent agreement as illustrated in Fig.~\ref{F:onetoone}(a).
Dispersion points for 1.7~T~$\parallel b^*$ were not included in those fits because this field, which is only just above the actual critical field $B_{\rm C}$, is below the classical transition field calculated to be $B_{\rm{C}}^{\rm cl}=1.72$~T for the best-fit parameters. At this field, quantum renormalization effects on the magnon dispersions are expected to be important, as discussed in a related context in Ref.~\cite{Gallegos2024}, as the zero-point quantum fluctuations are substantial close to $B_{\rm C}$. This is because the Hamiltonian $\mathcal{H}_{\rm XXZ}+\mathcal{H}_{\rm{Zeeman}}$ in Eqs.~(\ref{E:XXZnbcpo}) and (\ref{E:Zeeman}) does not have continuous rotational symmetry in finite transverse field $B_y\neq 0$, so the magnetization along the applied field direction is not a conserved quantity. It was empirically found that the observed dispersion at 1.7~T could be approximately captured by fixing the Hamiltonian parameters to those obtained from the fits to the higher-field data, and adding a variable effective field offset $\delta B$, i.e., in the Zeeman Hamiltonian in Eq.~(\ref{E:Zeeman}), $B_y$ was replaced by $B_{y}+\delta B$. We regard $\delta B$ not as an actual physical magnetic field, but as an empirical parameter to capture the effects of quantum fluctuations, which change the energy balance between the ordered and the polarized phase to make the polarized phase more stable. This empirical parameterization is only meaningful for fields in the range [$B_{\rm{C}}$, $B_{\rm{C}}^{\rm cl}$] to stabilize the paramagnetic field-polarized state at the mean field level. The effect of $\delta B$ on the dispersion is of a non-uniform increase of the magnon energies across the full band, larger for the bottom of the band and smaller for the top of the band. 
The best-fit Hamiltonian parameters are listed in Table~\ref{T:hamparams} and the level of agreement between the observed and calculated magnon energies is illustrated in Fig.~\ref{F:onetoone}(a). The excellent agreement is also illustrated in Fig.~\ref{F:highfieldresults}, which shows the best-fit linear spinwave theory dispersions overlain as white lines on top of the inelastic neutron scattering data. We have also compared the data with a Heisenberg-Kitaev model and concluded that this model can be ruled out; see Fig.~\ref{F:onetoone}(b) and Appendix~\ref{A:JKfits} for details.       

\begin{table}
\centering
\caption{Best-fit XXZ Hamiltonian parameters defined in Eqs.~(\ref{E:XXZnbcpo}) and (\ref{E:Zeeman}) with the level of agreement for magnon energies illustrated in Fig.~\ref{F:onetoone}(a). $\delta B$ is an empirical parameter to parameterize the dispersion renormalization effects at 1.7~T immediately above $B_{\rm C}$, as described in the text. 
}\begin{tabular}{cd}
\hline
\hline
$J_z$  & 0.1225(10)~\text{meV}\\
$J_{xy}$ & 0.0779(7)~\text{meV}\\
$g_c$ & 4.716(7) \\
$g_{ab}$ & 4.200(7)\\
$\delta B$ & 0.041(4)~\text{T at $B=1.7$~T }\parallel b^*\\
\hline
\hline
\end{tabular}
\label{T:hamparams}
\end{table}

\section{Excitations below $B_{\rm{C}}$}\label{S:lowINS}

\begin{figure*}
\includegraphics[width=\textwidth]{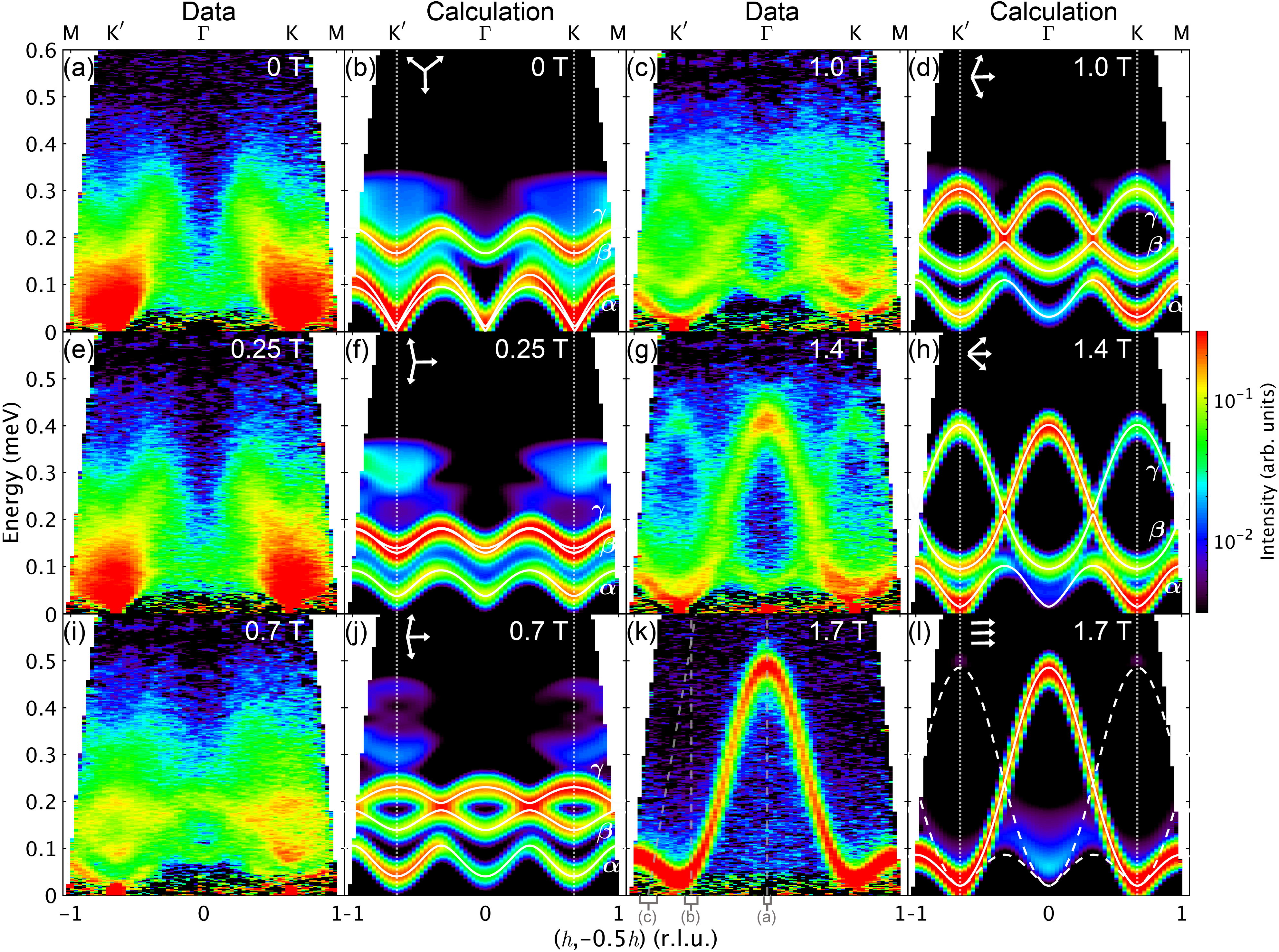}
\caption{(odd columns) INS data showing the evolution of the spectrum as a function of field~$\parallel b^*$, increasing from top to bottom and left to right. 
In each panel, the data are shown for in-plane wave vector along the $(1,-0.5)$ direction, and are averaged over a range in the $(0k0)$ direction that increases as a function of energy transfer (from $|k|<0.035$ on the elastic line to $|k|<0.055$ at $E=0.6$~meV) and over the full range of the data in the $(00l)$ direction, approximately $-1.3<l<0.5$. The data were taken with $E_i=0.87$~meV. (even columns) Calculations of the INS intensity in LSWT including both one- and two-magnon contributions. White solid lines show the corresponding magnon dispersions while white vertical dotted lines show locations of boundaries of the 2D Brillouin zone. The calculations use the classical 3-sublattice order at each field [see Fig.~\ref{F:phasediagram}(c)] illustrated by the three white arrows in the top left of the panels. 
For the zero field calculation in (b), a very small ($3\times10^{-3}$~T) notional magnetic field along $z$ was added to stabilize a Y phase (as was done for another Ising-like TLAF in Ref.~\cite{zhu2024continuumexcitationsspinsupersolidtriangular}) and the dynamical response was averaged over Y states related by arbitrary rotations around the $z$-axis. In (l) as 1.7~T is below the classical transition field $B_{\rm C}^{\rm cl}$, the effective field offset $\delta B$ was included in the calculation as per Table~\ref{T:hamparams}; the dashed white lines are the magnon dispersion shifted in wave vector by $\pm\bm{q}$, where $\bm{q}=(1/3,1/3)$ is the 2D propagation vector of the 3-sublattice magnetic order below $B_{\rm{C}}$. The intensity map in each calculation panel was obtained from the calculated dynamical response for every pixel contributing to the corresponding data panel, including the neutron polarization factor and isotropic \ch{Co^{2+}} magnetic form factor. Instrumental resolution effects are approximated by convolution with a Gaussian in energy of constant FWHM of 0.032~meV and averaging the response over the measured sample mosaic using a Monte-Carlo method. A common overall intensity scale factor was used in the calculations at all fields, obtained from a fit to the observed one-magnon intensity at the $\Gamma$ point in the 1.7~T data [Fig.~\ref{F:linewidths}(a) top trace]. The one- and two-magnon contributions at each field have been normalized to satisfy the sum rules as described in Appendix~\ref{A:2M}. In all panels, color represents scattering intensity on an arbitrary scale, with a log scale used in order for the continuum scattering to be visible in the calculation panels, and with an estimate of the non-magnetic background subtracted off in the data panels. In (k) the dashed gray lines over the data labeled as `(a)', `(b)' and `(c)' at the bottom of the panel indicate the location of the energy scans (rectangular brackets indicate the averaging range around the nominal momentum in the scan) in the corresponding panels in Fig.~\ref{F:linewidths}.
Note the slight left-right ($-h$ vs. $+h$) asymmetry in the intensity (but not the dispersions) in the data and calculations, which is due to the different $l$-ranges that the data is averaged over on the two sides, and subsequently different neutron polarization factors. 
}\label{F:twomagnon}
\end{figure*}

\subsection{Theoretical phase diagram in transverse field}\label{S:phasediagram}

Before discussing the evolution of the spectrum below $B_{\rm{C}}$, we briefly review the expected mean-field phase diagram in transverse field, illustrated in Fig.~\ref{F:phasediagram}(c). In zero field (solid dot) there is a continuous manifold of degenerate 3-sublattice ground states coplanar with the Ising axis \cite{Miyashita1985,Kleine1992}, from which an infinitesimally small transverse field stabilizes the `inverted Y' (iY) phase, in which one sublattice points along the field and the other two have one spin component opposite to the field, and the other component parallel to the Ising axis and opposite to each other. Upon increasing the field, the sublattices pointing away from the field progressively rotate towards the field, forming a $\Psi$ shaped structure, with no phase transition up until the transition to the field polarized (paramagnetic) `P' phase at the classical critical field $B_{\rm{C}}^{\rm cl}$. 

The phase diagram for the quantum $S=1/2$ model is expected to show some differences \cite{Gao2022aa,Yamamoto2019}, illustrated in Fig.~\ref{F:phasediagram}(b). In zero field, quantum fluctuations select the Y-phase from the degenerate manifold of classical states, resulting in two sublattices rotated symmetrically away from the Ising axis and the third sublattice pointing in the opposite sense along the Ising direction. This is expected to be stable up to a small transverse field $B_1$, above which a transition occurs to the same iY phase predicted semi-classically.  
Upon further increasing field, a further transition is predicted at $B_2$ to a V phase, stable up to the transition at $B_{\rm C}$ to the polarized phase. This V phase, which does not appear semi-classically, has three sublattices out of which two are parallel and form a honeycomb arrangement.

\subsection{Broadened magnons and excitation continua}\label{S:lowINSexpcalc}
In stark contrast to the sharp magnons observed above $B_{\rm C}$, the spectrum at 1.4~T and at all lower fields exhibits substantial continuum scattering and broadened excitations, as illustrated in Fig.~\ref{F:twomagnon} (odd columns). In a spinwave approach, a continuum of excitations described physically in terms of two-magnon scattering processes is expected generically to occur when the ground state has finite zero-point quantum fluctuations, which is indeed the case here at all probed fields. In order to gain understanding of the observed continuum scattering, the calculations shown in Fig. \ref{F:twomagnon} (even columns) therefore contain both one- and two-magnon contributions to the scattering intensity. 
Technical details of the one- and two-magnon calculations are given in Appendix~\ref{A:2M}. 

\begin{figure*}[htb]
\includegraphics[width=0.9\textwidth]{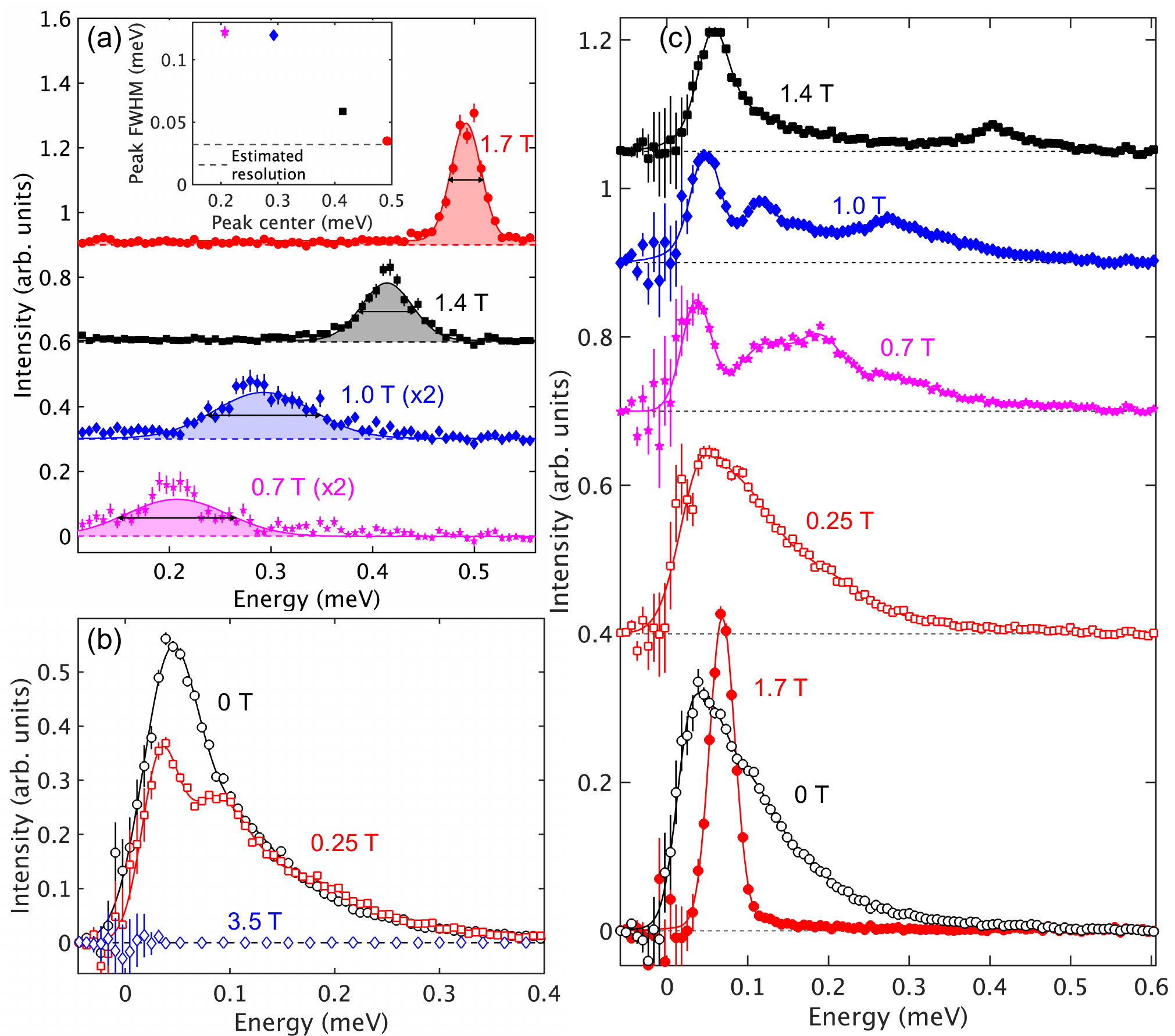}
\caption{Energy scans through the data in Fig.~\ref{F:twomagnon} as a function of field at three locations indicated by the gray dashed lines in Fig.~\ref{F:twomagnon}(k). (a) Energy scans at the $\Gamma$ point $(h,k)=(0,0)$ showing the increasing magnon linewidth with decreasing field. Data were averaged across $|h|, |k| <0.025$ in the $(h,-0.5h,0)$ and $(0k0)$ directions and $-1.2<l<0.3$ in the $(00l)$ direction. Solid lines and shaded areas show fits to Gaussian peak shapes and horizontal solid double headed arrows indicate the corresponding FWHM. For clarity, the intensities at 0.7 and 1.0~T have been multiplied by a factor of 2.  Inset: Linewidth (FWHM) of the fitted Gaussian peaks as a function of peak energy. The symbols used are the same as in the main panel. The estimated experimental energy resolution (dashed line) used in the calculations in Fig.~\ref{F:twomagnon} (even columns) is shown for reference. 
(b)~Energy scans slightly offset from the K$'$ point at $(h,k)=(-0.57,0.285)$ at 0~T (black circles), 0.25~T (red squares) and in the high-field polarized phase at 3.5~T (blue diamonds). There is strong continuum scattering across a significant energy range at 0~T and 0.25~T, but the data at 0.25~T have additional structure with a second broad hump around 0.1~meV. The data at 3.5~T are shown to illustrate the quality of the background subtraction. 
The data were averaged across $-0.62<h<-0.52$ in the $(h,-0.5h,0)$ direction, $|k|<0.055$ in the $(0k0)$ direction and the whole range of data in the $(00l)$ direction except on the elastic line where regions close to the magnetic Bragg peaks were masked out. (c) A combined energy-momentum scan along the nominal line $E=(5/3)(h_0+0.895)$, chosen so as to intersect almost transversely the dispersion relation of the lowest-energy quadratic mode near K$'$ for field in the range 0.7 to 1.4~T, to see this mode clearly separated from the higher-energy scattering signal. The lowest energy mode remains sharp down to 0.7~T. The data were averaged over $|h-h_0|<0.065$ in the $(h,-0.5h,0)$ direction, $|k|<0.055$ in the $(0k0)$ direction and the whole range of data in the $(00l)$ direction. In (b) and (c), solid lines are guides to the eye. In (a) and (c), curves have been offset vertically as a function of increasing field; dashed horizontal lines show the offset in each case. In all panels, $E_i=0.87$~meV and data points are neutron scattering intensities in arbitrary units with an estimate of the non-magnetic background subtracted off.
}\label{F:linewidths}
\end{figure*}

\subsubsection{Intermediate field phase}

As the transition to spontaneous magnetic order below $B_{\rm{C}}$ is continuous, one can qualitatively understand the evolution of the spectrum in the ordered phase immediately below the transition by applying Brillouin zone folding to the spectrum just above. 
Fig.~\ref{F:twomagnon}(l) shows the predicted spectrum at 1.7~T just above $B_{\rm{C}}$, where the solid white curve is the single magnon dispersion in Eq.~(\ref{E:lswtB}). The calculation captures the sharp magnon in the data well [Fig.~\ref{F:twomagnon}(k), which shows the same information as Fig.~\ref{F:highfieldresults}(a) but in a higher resolution configuration], and also predicts a small two-magnon contribution at low energies near $\Gamma$, which may well be below the sensitivity level of the present experiment. The magnetic structure just below $B_C$ is expected to have a three site unit cell, both in the quantum as well as in the mean-field phase diagrams for the XXZ TLAF Hamiltonian, although this magnetic structure takes different forms [see Figs.~\ref{F:phasediagram}(b) and (c), respectively], and so dashed white curves showing the shadow modes obtained by Brillouin zone folding associated with such a 3-sublattice magnetic order are also plotted. 

\begin{figure*}
    \includegraphics[width=0.9\textwidth]{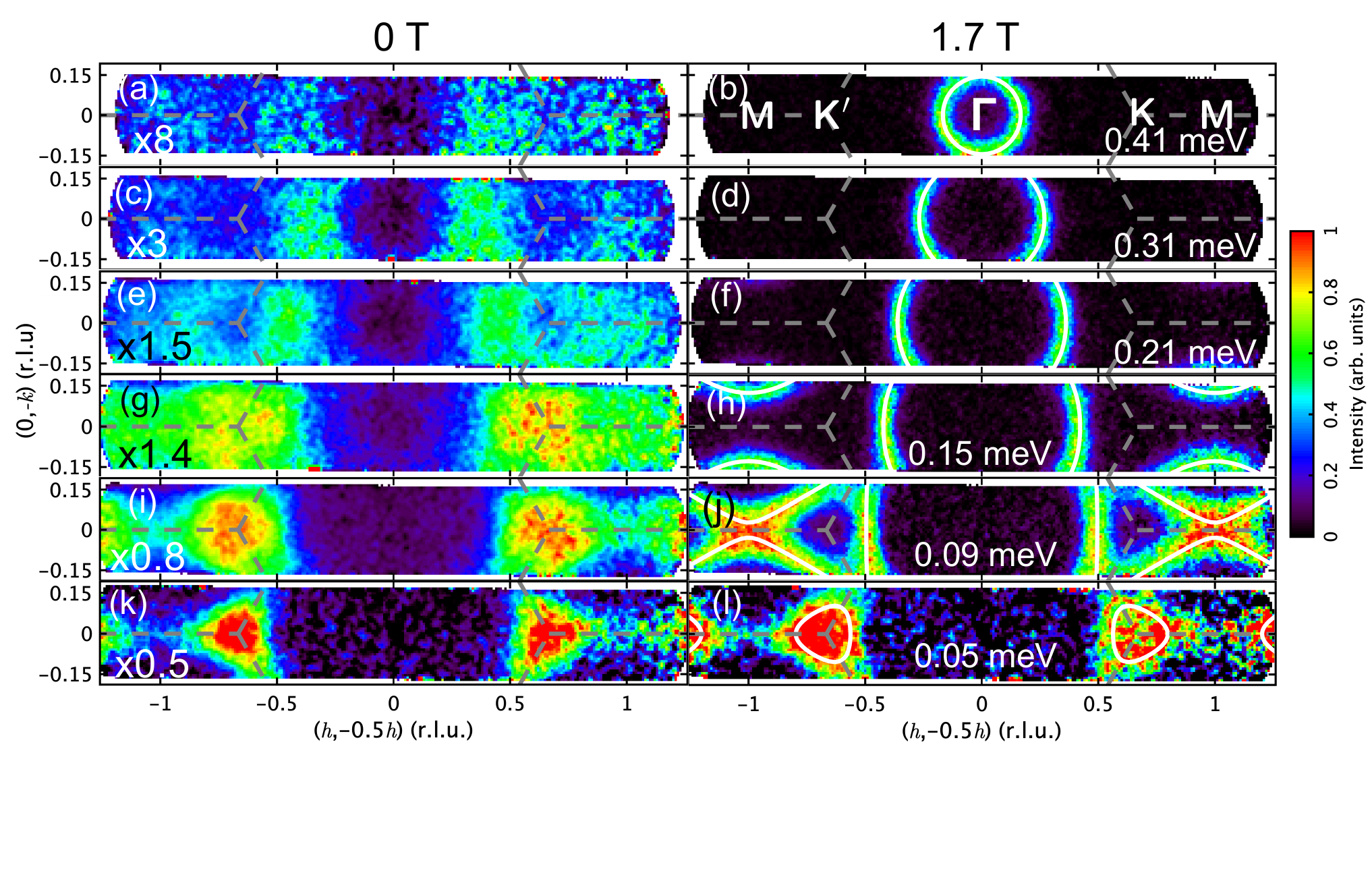}
    \caption{INS intensity at constant energy transfer as a function of two-dimensional momentum in the triangular planes in zero field (left) and in the field-polarized phase at 1.7~T (right) with energy increasing from bottom to top and indicated in the right-hand panels. The two-dimensional nature of the spectrum can be clearly seen, with a 3-fold rotational symmetry around the K and K$'$ points [see panels (j)-(l)] and near-circular symmetry for wave vectors near the $\Gamma$ point [see panel (b)]. Dashed gray lines indicate boundaries of the two-dimensional hexagonal Brillouin zones and white lines in the right-hand column are contours in momentum space of the calculated magnon dispersion in Eq.~(\ref{E:lswtB}) at the nominal energy of each panel, using the best-fit parameters in Table~\ref{T:hamparams}. In (b), high symmetry points of the two-dimensional Brillouin zone are indicated, showing the linear path MK$'\Gamma$KM used for the plots in Figs.~\ref{F:highfieldresults}, \ref{F:twomagnon}, \ref{F:degeneracy}(b) and \ref{F:Vphase}. The INS intensities were averaged over the whole wave-vector range in the $(001)$ direction and across a range of $\pm 0.01$~meV relative to the nominal energy transfer, as indicated by the labeled vertical black brackets on the left vertical axis in Fig.~\ref{F:highfieldresults}(a).  In all panels, color indicates the INS intensity on an arbitrary scale, 
from which an estimate of the non-magnetic background (including the incoherent scattering on the elastic line) has been subtracted off; the incident neutron energy was $E_i=1.41$~meV. The intensities in the left hand column only have been multiplied by the factor indicated in the bottom left corner of each panel.}
    \label{F:constantenergy}
\end{figure*}

Below $B_{\rm C}$, these shadow modes are expected to acquire finite intensity. In the calculated spectrum at 1.4~T in Fig.~\ref{F:twomagnon}(h), the three distinct magnon modes labeled $\alpha$, $\beta$, $\gamma$ in order of increasing energy evolve continuously from the solid and dashed lines in panel (l). The corresponding data at 1.4~T in panel (g) indeed show finite scattering weight at low energies near $\Gamma$ and high energies near K and K$'$, as predicted. The low-energy scattering near K$'$ also shows some incipient splitting [which becomes clearer at 1~T in panel (c)], again consistent with the theoretical prediction in panels (d) and (h). The gross features of the spectrum can thus be understood starting from the LSWT calculation. 

Distinct from the LSWT prediction, however, the magnon modes at intermediate to high energies appear visibly broadened and there is significantly more continuum scattering seen in the experiment than in the calculations. We attribute these effects to magnon decays due to interactions with this continuum of excitations, 
since strong magnon decays into magnon pairs are expected at all transverse fields below $B_{\rm{C}}$ due to the non-collinear magnetic structure \cite{Chernyshev2009,Zhitomirsky2013}.  
In contrast, the lowest energy modes near K and K$'$ appear to remain sharp; note for example in Fig.~\ref{F:twomagnon}(c) at 1~T the well-resolved quadratic mode at low energies around K$'$. The presence of a low-energy sharp magnon in this region is illustrated clearly in the energy scans in Fig.~\ref{F:linewidths}(c), where a sharp peak is present at the lowest energies in each of the top three traces showing data for 0.7 to 1.4~T. Note that, at 0.7 and 1~T, 
the sharp peak is clearly separated from a higher-energy scattering signal, which is in the form of broad peaks and/or continuum scattering, with the relative weight in the sharp mode decreasing upon decreasing field and more weight moving into the higher-energy broad continuum scattering. Energy scans at the $\Gamma$-point shown in Fig.~\ref{F:linewidths}(a) illustrate how the sharp magnon mode observed at 1.7~T (top trace) just above $B_{\rm C}$ gradually broadens upon lowering field (lower traces) and reduces its energy due to the reduction in Zeeman energy. The inset in Fig.~\ref{F:linewidths}(a) illustrates how the peak width increases as the magnon energy reduces upon lowering field: while at 1.7~T the magnon mode is resolution limited, at all fields below $B_{\rm C}$ the peak width far exceeds the estimated experimental energy resolution (dashed line in the inset). 

These observations suggest that the intermediate- ($\beta$) and higher-energy ($\gamma$) magnons do not survive as sharp modes in the ordered phase below $B_{\rm C}$. We have verified that it is indeed kinematically possible (for details see Appendix~\ref{A:2M}) for the $\beta$ mode to decay into two-magnon $\alpha\alpha$ states at all of 0.7, 1 and 1.4~T, and for the $\gamma$ mode to decay into $\beta\beta$, $\beta\gamma$ or $\alpha\beta$ continua at 1.4~T, the $\beta\beta$ or $\alpha\beta$ continua at 1~T, and the $\alpha\beta$ or $\alpha\alpha$ continua at 0.7~T. The $\alpha$ mode is expected to be kinematically forbidden from decaying throughout this field range, consistent with the observation that the lowest energy mode appears to remain sharp in this regime. 

\subsubsection{Low field phase}

Considering now lower fields, we note that a quantum calculation for the XXZ TLAF predicts a first order phase transition between an iY and a V phase \cite{Gao2022aa} [at a field $B_2$, see Fig.~\ref{F:phasediagram}(b)], and indeed thermodynamic measurements do indicate anomalies associated with a phase transition near 0.5~T \cite{Sheng2022}, as shown in Fig.~\ref{F:phasediagram}(a), so the spectrum is not expected to evolve continuously across this field. Indeed, the experimental data in Fig.~\ref{F:twomagnon}(a) at 0~T and (e) at 0.25~T look very different both to the data at higher field and to the calculated spinwave spectra in panels (b) and (f), respectively. The data have no detectable sharp modes or magnon-like features, contrary to what is predicted and such as are seen down to 0.7~T near the K$'$ point [Fig.~\ref{F:twomagnon}(i)], and the continuum in the calculations is both far weaker than what is seen experimentally and has a very different shape. The spectrum is not however featureless; the momentum and energy dependence of the zero field spectrum are studied in more detail in Fig.~\ref{F:constantenergy} (left column), revealing very broad but still dispersive features with a strongly 2-dimensional nature. Note in particular the triangular contours around K and K$'$ at the lowest energies in panel (k).  

For completeness, we note that even though the overviews of the spectra at 0 and 0.25~T plotted in Figs.~\ref{F:twomagnon}(a) and (e) appear very similar (with intensity on a log scale), there are differences in the detail of the lineshapes at low energies between the two fields. This is illustrated clearly in the energy scan in Fig.~\ref{F:linewidths}(b) for a wave vector slightly offset from K$'$. While at 0~T the lineshape has the form of a monotonic decrease with increasing energy, at 0.25~T the intensity at the lowest energies is suppressed and there appears to be more structure in the form of a second broad hump at the intermediate energy of 0.1~meV.   

In terms of magnon decay effects, we have verified that, in zero field for the $\gamma$ mode, there is a large density of states for decay into the $\beta\beta$, $\alpha\alpha$ or $\alpha\beta$ continua. The $\beta$ mode also overlaps with the $\alpha\alpha$ continuum in a region where the latter has a small, but finite density of states, so this may also provide decay channels. At 0.25 T, the $\beta$ and $\gamma$ modes both have a large density of states to decay into the $\alpha\alpha$ continuum. On the other hand, the $\alpha$ mode has no kinematically allowed decays at either 0 or 0.25 T, so would be expected to appear as a sharp, prominent dispersive feature, which, however, is clearly not observed in the data.

\subsection{Classical ground-state degeneracy}\label{S:degeneracy}

Another aspect that could be relevant for understanding the observed spectrum of excitations in zero field is the fact that the Ising-like XXZ TLAF model has a continuous manifold of mean-field degenerate ground states, already mentioned when discussing the magnetic phase diagram in Sec.~\ref{S:phasediagram}. 
\begin{figure*}[tb]
\includegraphics[width=\linewidth]{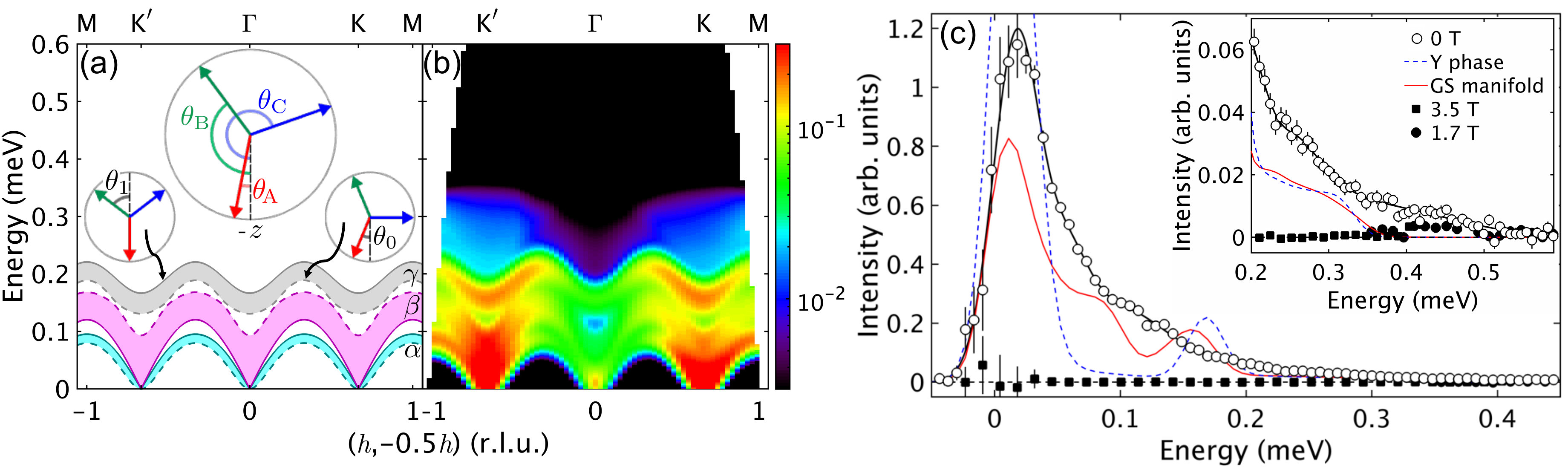}
\caption[Manifold of degenerate mean-field states for the XXZ Ising model]{(a) (top) Schematic of the manifold of mean-field degenerate ground states for the Ising-like XXZ TLAF Hamiltonian in Eq.~(\ref{E:XXZnbcpo}) parameterized by Eq.~(\ref{E:mfangles}). Colored arrows are the spins of the three sublattices and all indicated angles are measured in a clockwise sense from the $-z$ axis. (main panel) Dispersion relations vary between those different ground states as indicated by the shaded regions, bounded by the curves for the Y phase (solid lines, left sketch, $\theta_{\rm A}=0$, $\theta_{\rm B,C}=\pi\mp\theta_1$), and the transverse-Y phase (dashed lines, right sketch, $\theta_{\rm A}=\theta_0$, $\theta_{\rm B}=\pi-\theta_0$, $\theta_{\rm C}=3\pi/2$). 
The different color shading corresponds to the three distinct spinwave modes $\alpha$, $\beta$ and $\gamma$. (b) Calculated INS spectrum including both one- and two-magnon excitations, averaged over the full manifold of mean-field degenerate ground states, i.e., the family of states shown in the sketches in (a) as well as those obtained from those structures by an arbitrary rotation of all spins around $z$. The calculation includes all experimentally relevant factors --- including the same scale factor --- as in the calculations for the Y phase in Fig.~\ref{F:twomagnon}(b), and are to be compared directly with the INS data in Fig.~\ref{F:twomagnon}(a). (c) Observed INS intensity in zero field  (open circles) in an energy scan at the Brillouin zone corner point K$^\prime$(-2/3,1/3) with a black solid line as a guide-to-the-eye, compared with the calculation in (b) (red solid line), and calculation for a single ground state (blue dashed line) for the Y phase as per parameters in Fig.~\ref{F:twomagnon}(b). The inset is a zoomed-in view of the high-energy tail. The transverse momentum averaging ranges along $(h,-0.5h,0)$ and $(0k0)$ vary from $|h+2/3|,|k|<0.035$ on the elastic line to $|h+2/3|<0.064$, $|k|<0.055$ at $E=0.6$~meV, with the full range of $l$ included, except on the elastic line where the wave-vector regions near the 3D magnetic Bragg peaks are masked out so as to to obtain the purely inelastic magnetic signal. An estimate of the non-magnetic background has been subtracted off; the quality of the background subtraction is illustrated by the data at 1.7 and 3.5~T (filled circles and squares, respectively), for which no intrinsic magnetic scattering is expected in the range of the plotted data. 
}\label{F:degeneracy}
\end{figure*}
All these states have three sublattices with spins coplanar with the Ising axis and making different angles $\theta_{\rm A}$, $\theta_{\rm B}$ and $\theta_{\rm C}$ with the $-z$ axis, as illustrated in Fig.~\ref{F:degeneracy}(a) (top inset). The Y-phase quantum ground state, which has been considered so far, is one of these states, and has the A spin down ($\theta_{\rm A}=0$) and the B and C spins up-left and up-right at $\theta_{\rm B,C}=\pi\mp\theta_1$ [illustrated in Fig.~\ref{F:degeneracy}(a) (left circle)] with \cite{Miyashita1985} 
\begin{equation}
\theta_1=\arccos\frac{1}{1+\Delta}, \quad \Delta=\frac{J_{xy}}{J_z}. \nonumber
\end{equation}
The degeneracy arises because the condition for minimal energy imposes only two constraints on the three angles, so $\theta_{\rm B,C}$ can be regarded as functions of $\theta_{\rm A}$, which can vary continuously. To generate symmetry-distinct magnetic structures it is sufficient to vary $\theta_{\rm A}$ in the range $[0,\theta_0]$, where the lower limit gives the Y phase, and the upper limit \cite{Kleine1992} 
\begin{equation}
\theta_0=\arcsin\frac{\Delta}{1+\Delta},
\label{E:theta0}
\end{equation}
gives the transverse-Y structure illustrated in Fig.~\ref{F:degeneracy}(a) (right circle) with the C spin horizontal.  
For an intermediate $\theta_{\rm A}$, one obtains analytically $\theta_{\rm B,C}=\pi+\epsilon\mp\delta$ \cite{Kleine1992},       
with\footnote{We have corrected a typo in the last term in the denominator of these equations $\Delta \cos^2 \theta_{\rm{A}} \rightarrow \Delta^2 \cos^2 \theta_{\rm{A}}$ from Ref.~\cite{Kleine1992} to get the correct expression for $\theta_0$ in Eq.~(\ref{E:theta0}). We have also verified numerically using SpinW \cite{Toth2015nu} that the above equations give a mean-field energy (per spin) \cite{Kleine1992} $e_{\rm cl}/S^2=-J_z(1+\Delta+\Delta^2)/(1+\Delta)$ independent of $\theta_{\rm{A}}$.} 
\begin{eqnarray}
\cos \delta & = & \frac{\Delta}{(1+\Delta){(\sin^2\theta_{\rm A}+\Delta^2 \cos^2\theta_{\rm A}})^{1/2}}, \nonumber\\
\cos \epsilon & = & \frac{\Delta\cos\theta_{\rm A}}{(\sin^2\theta_{\rm A}+\Delta^2 \cos^2\theta_{\rm A})^{1/2}},\label{E:mfangles}
\end{eqnarray}
and with $\theta_{\rm B}\in[\pi-\theta_1,\pi-\theta_0]$ and $\theta_{\rm C}\in[\pi+\theta_1,3\pi/2]$. Within this family of states, all three angles vary monotonically in the same sense, but at different rates such that the relative angles between the three spins change between the different structures. The mean-field degeneracy of this family of states is inherited from the $\Delta=1$ Heisenberg limit when the above ground states become 120$^\circ$ coplanar states related to one-another by rigid rotations, which are symmetry operations of the Hamiltonian. In the Ising-like case, there is no rotational symmetry normal to the $z$-axis, but the above states are related by a pseudo-symmetry that remains at the mean-field level. Those different structures have slightly different dispersion relations as indicated in Fig.~\ref{F:degeneracy}(a) by the shaded areas, bounded by the dispersions (solid/dashed lines) in the extreme cases of a Y-phase or a transverse-Y phase illustrated in the left/right diagrams, respectively. 

The continuous manifold of mean-field degenerate ground states could potentially be relevant for understanding the observed zero-field spectrum in \ch{Na2BaCo(PO4)2} if the magnetic propagation vector between layers is incommensurate, which is one of the two generic scenarios for the 3D magnetic structure discussed in the following Sec.~\ref{S:interlayer}. In this scenario, frustrated interlayer couplings favor an incommensurate rotation angle between the spins of adjacent layers, and in a first approximation one can consider that the magnetic structure in each layer is close to one of the degenerate mean-field ground states of decoupled layers, with each such state occurring with the same frequency across the full stack of layers due to the incommensurate periodicity along $c$. Neglecting dynamical effects of the (weak) interlayer couplings, the INS spectrum of such a system would be obtained by averaging the spectrum uniformly over the layers, so over the full manifold of degenerate mean-field ground states, and the result is plotted in Fig.~\ref{F:degeneracy}(b). A similar averaging over degenerate ground states was recently considered in the XXZ TLAF material \ch{CeMgAl11O19}, which has Hamiltonian parameters close to the special ratio $J_z/J_{xy}=-0.5$ where a continuous degeneracy of umbrella ground states is expected~\cite{gao2024spinexcitationcontinuumexactly}. 

The spectrum averaged over degenerate ground states in Fig.~\ref{F:degeneracy}(b) is clearly broadened compared to that of the (single) Y ground state illustrated in Fig.~\ref{F:twomagnon}(b), which brings the model closer to the data in Fig.~\ref{F:twomagnon}(a), although notable discrepancies still remain. Those are clearly illustrated in the energy scan at the zone corner K$^{\prime}$ plotted in Fig.~\ref{F:degeneracy}(c), where the solid red/dashed blue lines show calculations for the manifold average/single Y phase, respectively; both show considerably more structure than the data (open circles). Including magnon decay effects for the $\beta$ and $\gamma$ magnons would be expected to broaden further the predicted lineshape profile in the intermediate energy range and this could potentially improve the agreement with the data. However, it is unclear if such effects could ultimately fully explain the large and extended high-energy tail seen in the experimental data and highlighted in the inset of Fig.~\ref{F:degeneracy}(c), or whether a fundamentally different approach that aims to solve the full quantum problem, such as approaches based on tensor network states \cite{Sheng2024,gao2024spinsupersolid,Chi2024}, will be needed to fully explain the observed continuum-dominated spectrum in zero and low field.

\section{Long-range magnetic order and interlayer couplings}\label{S:diffraction}

\begin{figure}[htb]
\includegraphics[width=0.5\textwidth]{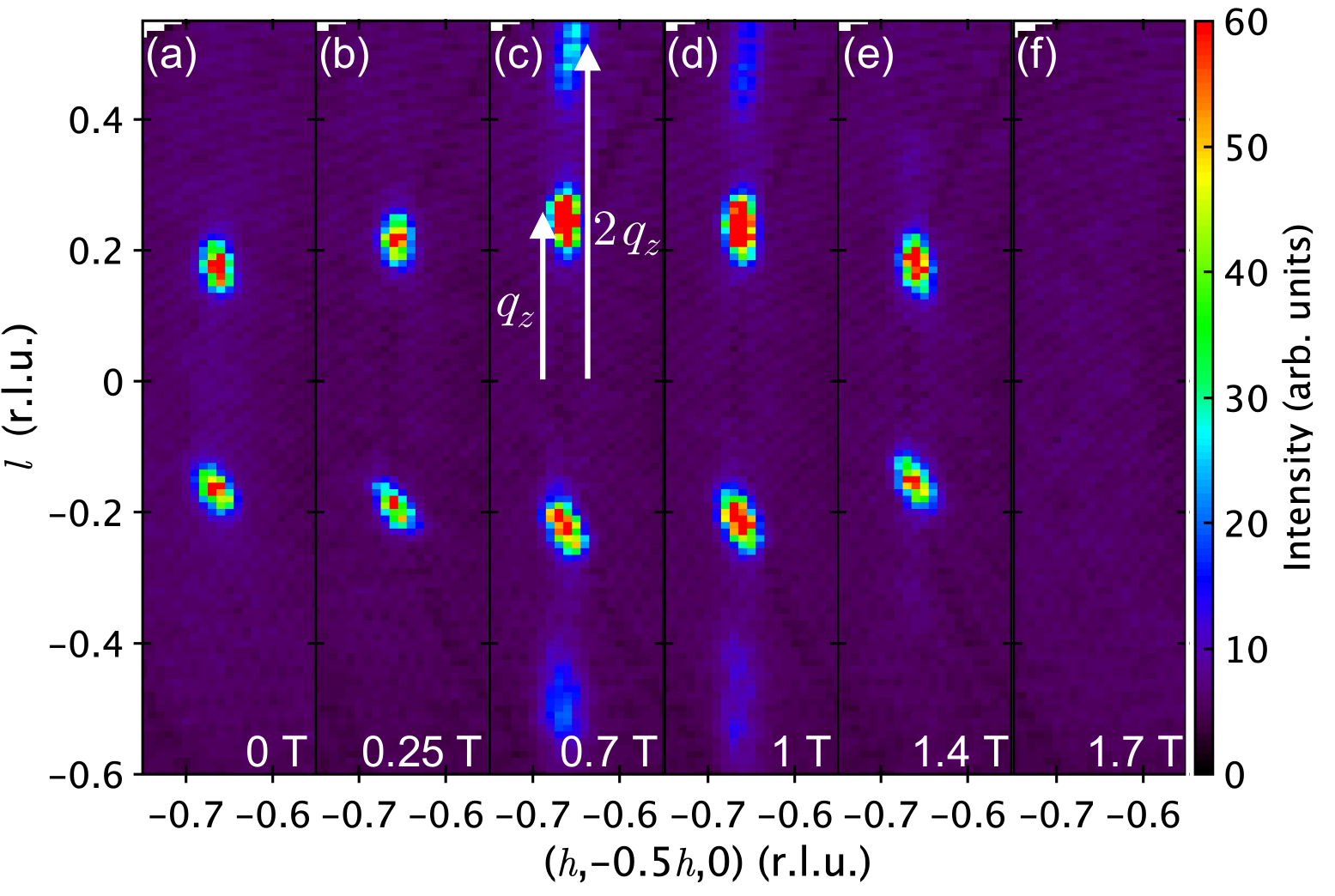}
\caption{Evolution of the scattering intensity on the elastic line as a function of increasing magnetic field from left to right. The sharp spots are Bragg peaks associated with spontaneous 3D magnetic order. In each panel, the horizontal axis is the wave-vector component along the in-plane $(h,-0.5h,0)$ direction and the vertical axis is the wave-vector component along the out-of-plane $(00l)$ direction; all above peaks occur at $h=-2/3$, indicating a 3-sublattice in-plane magnetic ordering. In panels (c) and (d), weaker peaks are also observed at the second harmonic $\pm 2q_z$ positions in addition to the primary peaks at $\pm q_z$, where $q_z$ is the component of the magnetic propagation vector along the interlayer direction. In all panels, color shows the raw neutron scattering intensity on an arbitrary scale. The data have been averaged across $|k|<0.05$ in the $(0k0)$ direction and over $|E|<0.02$ where $E$ is the energy transfer in meV. The incident energy was $E_i=0.87$~meV.
}\label{F:elastic_line}
\end{figure}

\begin{figure}[htb]
\includegraphics[width=0.3\textwidth]{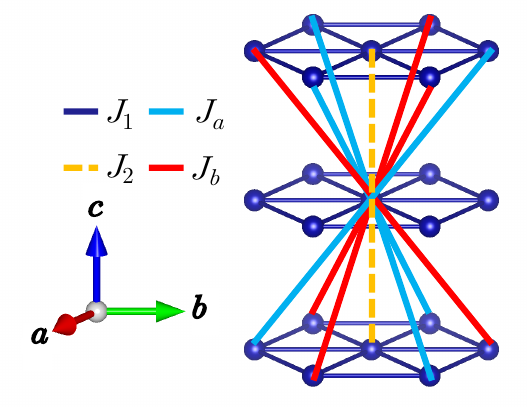}
\caption{Schematic diagram of interlayer exchange paths. The $J_a[1,0,1]$ (light blue) and $J_b[1,1,1]$ (red) bonds are symmetry-equivalent in the nominal $P\bar{3}m1$ structure, but are symmetry-inequivalent in the $P\bar{3}$ structure refined in Appendix~\ref{A:xrays}. 
}\label{F:interlayer_interactions}
\end{figure}

\subsection{Interlayer couplings}\label{S:interlayer}

In zero field, we observe sharp magnetic Bragg peaks on the elastic line [see Fig.~\ref{F:elastic_line}(a)] at positions associated with magnetic propagation vectors $(1/3,1/3,\pm q_z)$, with $q_z$ consistent, within the available wave-vector resolution, with the range of values 0.167(3) to 0.183(11) quoted in previous studies \cite{gao2024spinsupersolid,Xiang2024aa,Sheng2022}. We note that magnetic propagation vectors of this form have been observed before for stacked magnetic triangular layers in RbFe(MoO$_4$)$_2$ ($q_z=0.44$) and attributed to frustrated interlayer couplings \cite{Hearmon2012}. We first discuss the key physics, which is illustrated by a simplified model with Heisenberg exchanges --- $J_1$ between nearest-neighbors in plane, and interlayer couplings $J_{2,a,b}$ defined in Fig.~\ref{F:interlayer_interactions} --- and later discuss the effects of XXZ exchange anisotropy on the $J_1$ bond. The magnetic structure in the Heisenberg case has the ordered spins rotate by $120^{\circ}$ between nearest-neighbors in the same layer and by $2\pi q_z$ between vertically-stacked sites in adjacent layers. The mean field energy per site of this magnetic structure is obtained analytically as\footnote{$J_{a,b}$ are defined to be a factor of 3 smaller than in Ref.~\cite{Hearmon2012}, such that $J_{a,b}$ are the exchange energies per bond.} \cite{Hearmon2012}
\begin{align}
\frac{E}{S^2}=-\frac{3}{2}J_1 -& \frac{1}{2}\big[3(J_a+J_b)-2J_2\big]\cos{2\pi q_z} \nonumber \\
-&\frac{3\sqrt{3}}{2}(J_b-J_a)\sin{2\pi q_z},
\label{E:en}
\end{align}
with the minimum energy achieved for $q_z$ given by 
\begin{equation}
\tan(2\pi q_z)=\frac{3\sqrt{3}(J_b-J_a)}{3(J_b+J_a)-2J_2}.
\label{E:qz}
\end{equation}
In the nominal $P\bar{3}m1$ crystal structure \cite{Zhong2019}, the $J_a$ and $J_b$ bonds are symmetry-related by vertical $m$ mirror planes. In that case, the stable configuration therefore corresponds to $q_z=0$ or $0.5$, i.e., ferromagnetic or antiferromagnetic stacking of layers, depending on whether $3J_b-J_2>0$ or $<0$. The intermediate value $q_z$ observed experimentally cannot be explained within this framework. However, if the crystal structure has subtle distortions that break the vertical mirror planes and reduce the space group symmetry to $P\bar{3}$ as in RbFe(MoO$_4$)$_2$, the exchange paths $J_a$ and $J_b$ become symmetry-distinct and in this case Eq.~(\ref{E:qz}) gives an intermediate rotation angle of the ordered spins between layers. We provide evidence in Appendix~\ref{A:xrays} that such a distortion does indeed occur in the actual crystal structure. Note that swapping $J_a$ and $J_b$ in the above two equations, equivalent to rotating the crystal by 180$^\circ$ around the $\bm{a}+\bm{b}$ axis in Fig.~\ref{F:interlayer_interactions}, changes the sign of $q_z$, so without loss of generality it is sufficient to discuss only the case $0<q_z<0.5$ realized for $J_b>J_a$. In this case $q_z<0.25$ or $>0.25$ depending on whether $3(J_b+J_a)-2J_2>0$ or $<0$. In the special case $J_{2,a}=0$ and $J_b>0$, the minimal energy is obtained for $q_z = 1/6$, which is within the range of values seen experimentally. For simplicity, this case is assumed in Appendix~\ref{A:NBCPO_interlayer} to estimate an upper bound on the strength of the interlayer couplings, as there is not enough information to distinguish between the different possibilities. We note that starting from this special case and adding $J_2>0$ and/or $J_a<0$ would increase $q_z$ and make it incommensurate; for a given incommensurate $q_z$ there is, in principle, a whole family of possible combinations of $J_{2,a,b}$. 

Interestingly, a magnetic propagation vector of the same form with $q_z=0.1869(3)$ was reported in the spin-5/2 analogue \ch{Na2BaMn(PO4)2} \cite{Zhang2024aa}, in which the exchange interactions are expected to be near-Heisenberg. The easy-axis spin-1 \ch{Na2BaNi(PO4)2} also has the same form of propagation vector, but with a larger $q_z=0.293(1)$ \cite{Sheng2025aa}. Both of these materials have also been found to exhibit a crystal symmetry lowering from $P\bar{3}m1$ to $P\bar{3}$ \cite{Kajita2024aa}, suggesting a general applicability of the minimal model in Fig.~\ref{F:interlayer_interactions} to explain the finite interlayer magnetic propagation vector.

\begin{figure}[htb]
\includegraphics[width=0.32\textwidth]{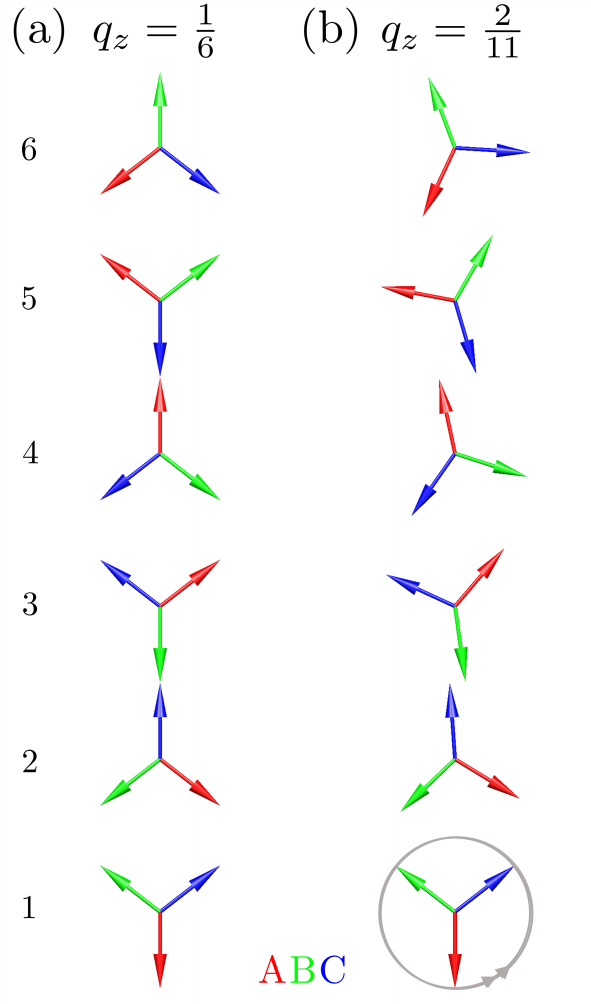}
\caption{Magnetic structure models for stacked layers with an XXZ Hamiltonian and interlayer paths as in Fig.~{\ref{F:interlayer_interactions}}. Colored arrows show the spin orientations at sites A (red), B (green) and C (blue), displaced vertically according to the layer index shown on the left. The B and C sites are offset in-plane relative to A by $\bm{a}$ and $\bm{a}+\bm{b}$, respectively. (a) A 6-layer periodic structure obtained by starting from the Y-phase in the basal layer and successively applying translation by $\bm{a}+\bm{b}+\bm{c}$ combined with time reversal. The B sites in each layer are antiparallel to A sites in the layer above and to C sites in the layer below. This structure is stabilized by $J_b>0$ and $J_{2,a}=0$. (b) A representative magnetic structure with $q_z\neq1/6$, where the spin arrangements in adjacent layers are {\em not} related by symmetry operations, but involve in general incommensurate angle rotations between spins in successive layers. The double arrows on the circular envelope for the basal layer indicates the sense of spin rotation at each of the three sites A, B and C upon moving up between layers. Such a structure is stabilized starting from the model in (a) by switching on finite $J_2>0$ or $J_a<0$, as described in the text. 
}\label{F:mag_structure}
\end{figure}

In order to explore further the applicability of this framework to the experimentally-relevant case of substantial Ising-like anisotropy of the $J_1$ bond, we used the numerical ground state energy minimization in SpinW~\cite{Toth2015nu} to determine the mean-field ground state for the XXZ spin Hamiltonian in Eq.~(\ref{E:XXZnbcpo}) and Table~\ref{T:hamparams} for various interlayer exchange models. In order to capture boundary effects accurately, we considered commensurate magnetic unit cells, with propagation vector in the vicinity of the experimentally-observed range of $q_z$ values. Two qualitatively different magnetic structures are illustrated in Fig.~\ref{F:mag_structure}. Panel (a) shows a structure with $q_z=1/6$, obtained from one of the three-sublattice degenerate ground states of the XXZ model (chosen here as the Y phase) and successively applying translation by $\bm{a}+\bm{b}+\bm{c}$ combined with time reversal. This structure is stable for $J_b>0$ and $J_{2,a}=0$, with all $J_b$ bonds fully satisfied. Adding finite $J_2>0$ and/or $J_a<0$ frustrates this spin structure and favors instead a structure with a larger, and in general incommensurate, $q_z$. An example of such a structure is shown in panel (b) for $J_a=0$, $J_b=0.003$~meV (as estimated in Appendix~\ref{A:NBCPO_interlayer}) and $J_2/J_b=1/3$ chosen to stabilize $q_z=2/11$. 
The main qualitative difference between the (a) and (b) structures is that in (a) the spin arrangement in every layer is the same up to symmetry operations (integer in-plane lattice translations and time reversal), whereas in (b) the spin arrangement in adjacent layers cannot be related by symmetry operations. Which type of magnetic structure is ultimately realized in \ch{Na2BaCo(PO4)2} will depend on the fine balance between quantum fluctuations that prefer a Y phase in each layer and the strengths of the interlayer couplings that might favor spin rotation between layers, and possibly other interactions and anisotropies beyond the minimal model described here.         

\subsection{Evolution of the magnetic Bragg peaks with transverse field}\label{S:diffractionB}
For completeness, Fig.~\ref{F:elastic_line} shows an overview of the evolution of the magnetic diffraction pattern as a function of increasing applied transverse field. The magnetic propagation vectors remain of the form $(1/3,1/3,\pm q_z)$ at all fields, but with a field-dependent $q_z$. At 0.25~T, $q_z$ has increased only slightly compared to the zero field value, whereas at 0.7~T, after the transition to what is believed to be the V phase \cite{Gao2022aa}, the diffraction pattern shows a qualitative change. Here, $q_z$ is seen to increase further to nearly $1/4$, and a second set of weaker diffraction peaks appear at the satellite position $\pm 2q_z$, i.e., these peaks are the second harmonics of the main magnetic satellites at vertical wave-vector components $\pm q_z$ [see Fig.~\ref{F:elastic_line}(c)]. These second harmonic peaks indicate that a single Fourier component cannot fully capture the magnetic structure in this intermediate field regime and they are suppressed faster than the main satellites upon increasing field [see Fig.~\ref{F:elastic_line}(c)-(e)]. As the field increases above 0.7~T, $q_z$ decreases again towards $1/6$ and at 1.7~T [panel (f)] all magnetic Bragg peaks have disappeared, confirming the transition to the high-field polarized phase. The evolution of the $l$-dependence of the magnetic diffraction pattern upon increasing field is intriguing, but an understanding of this effect is beyond the scope of the present work.

\section{Conclusions}\label{S:conclusion}
In summary, we have reported high resolution inelastic neutron scattering measurements of the spin excitations in single crystals of the Ising-like TLAF \ch{Na2BaCo(PO4)2} in magnetic field applied transverse to the Ising axis. In the polarized phase at high field, sharp magnons are observed, which are well described by linear spinwave theory for a nearest-neighbor Ising-like XXZ exchange model. Below the critical polarizing field, however, strong continuum scattering is seen, much stronger than expected semi-classically by linear spinwave theory. While the spectrum down to 0.7~T could be plausibly described by magnons strongly broadened at intermediate and high energies by interactions with two-magnon excitation continua, the spectrum at 0.25~T and 0~T is dominated by a strong scattering continuum with no sharp features at any of the energies probed, in spite of the presence of 3D magnetic Bragg peaks confirming long-range magnetic order. We discussed the possible relevance, for capturing the experimentally-observed spectrum in zero field, of the presence of a continuous manifold of mean-field degenerate ground states and we provided a quantitative comparison of the data with the average spectrum of such a manifold. We hope our quantitative analysis will stimulate the further development of concrete theoretical predictions for the possible role of magnon decay effects, or advance more sophisticated quantum models of the spin dynamics, which could be directly compared with the experimental data to gain further insight into the physics of the spin-1/2 antiferromagnetic triangular lattice Ising-like XXZ model. 
We also proposed models of interlayer couplings to explain the finite magnetic propagation vector in the interlayer direction, which requires the breaking of mirror planes in the nominal crystal structure. Through x-ray diffraction from small, untwinned crystals, we provided direct evidence for such a symmetry lowering and proposed a revised crystal structure model. 

Access to the data will be made available from Ref.~\cite{data_archive}.

\begin{acknowledgments}
L.W. and R.C. acknowledge useful discussions with Cristian Batista and Daniel Flavi\'{a}n Blasco. This research was partially supported by the European Research Council under the European Union’s Horizon 2020 research and innovation programme Grant Agreement Number 788814 (EQFT). L.W. acknowledges support from a doctoral studentship funded by Lincoln College, the University of Oxford and the ERC grant above, and from a grant from the UKRI International Science Partnerships Fund (award ISPF-229) for partnership development between ISIS, Diamond and the Paul Scherrer Institute. R.O. acknowledges support from JST ASPIRE (Grant No.~JPMJAP2314). The neutron scattering measurements at the ISIS Facility were supported by a beamtime allocation from the Science and Technology Facilities Council \cite{LET_doi1,LET_doi2}. 
Crystal structure figures were made using \textsc{vesta} \cite{Momma2011}.
\end{acknowledgments}

\appendix

\section{Revised crystal structure}\label{A:xrays}
Here we report single crystal x-ray diffraction measurements, which confirm the breaking of the mirror planes in the nominal $P\bar{3}m1$ crystal structure \cite{Zhong2019}, as required by the model of interlayer magnetic couplings proposed to explain the finite magnetic propagation vector along the interlayer direction in Sec.~\ref{S:interlayer}. Our studies are complementary to previous neutron diffraction studies on powder samples \cite{Kajita2024aa}, which also found evidence for such a crystal symmetry reduction. 

\begin{figure}[tbh]
\includegraphics[width=0.4\textwidth]{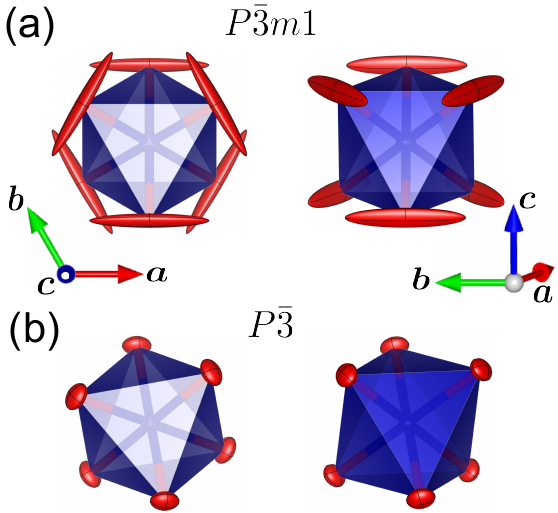}
\caption{View of the CoO$_6$ octahedron (blue shading) from different directions (left/right panels) in relation to the hexagonal crystal axes, obtained from refinement of single-crystal x-ray diffraction data at 100~K against different structural models. (a) Nominal $P\bar{3}m1$ structure with parameters in Table~{\ref{T:XRD_100K_P-3m1}}. (b) Lower-symmetry $P\bar{3}$ structure with parameters in Table~{\ref{T:XRD_100K_P-3}}. The red ellipsoids illustrate the anisotropic displacement parameters at the oxygen O1 positions, with black line contours along the principal planes. The refinement agreement in the two cases is illustrated in Figs.~{\ref{F:1to1}}(a) and (b), respectively. 
}\label{F:ADPs}
\end{figure}

Single-crystal x-ray diffraction data were collected using a Mo source SuperNova diffractometer both at room temperature and at 100~K under N$_2$ gas flow. The integrated diffraction intensities were corrected for absorption using numerical integration over a multifaceted crystal model within CrysAlis\textsc{PRO} \cite{CrysAlisPro} and structural refinement was performed using \textsc{Olex2} \cite{Olex2}. Refinement against the nominal $P\bar{3}m1$ structure using isotropic displacement parameters gave systematically unphysically large displacement parameters for the O1 oxygens coordinating the Co ions compared to all other atoms in the unit cell, an effect also noted in the analysis of early powder x-ray data in Ref.~\cite{Zhong2019}. A significant improvement in the refinement was obtained by allowing anisotropic displacement parameters (ADPs), which gave highly-elongated O1 oxygen ellipsoids in a direction in the $ab$ plane and normal to the Co-O bonds, as illustrated in Fig.~\ref{F:ADPs}(a) for Sample~I at 100~K with parameters listed in Table~{\ref{T:XRD_100K_P-3m1}}. These large oxygen ADPs did not change significantly upon cooling between room temperature and 100~K. In contrast, all the other atoms in the unit cell had relatively much smaller ADP parameters already at room temperature (not shown), which also decreased significantly (by about $\sim50\%$) upon cooling to 100~K.  

\begin{figure*}[tbh]
\includegraphics[width=0.95\textwidth]{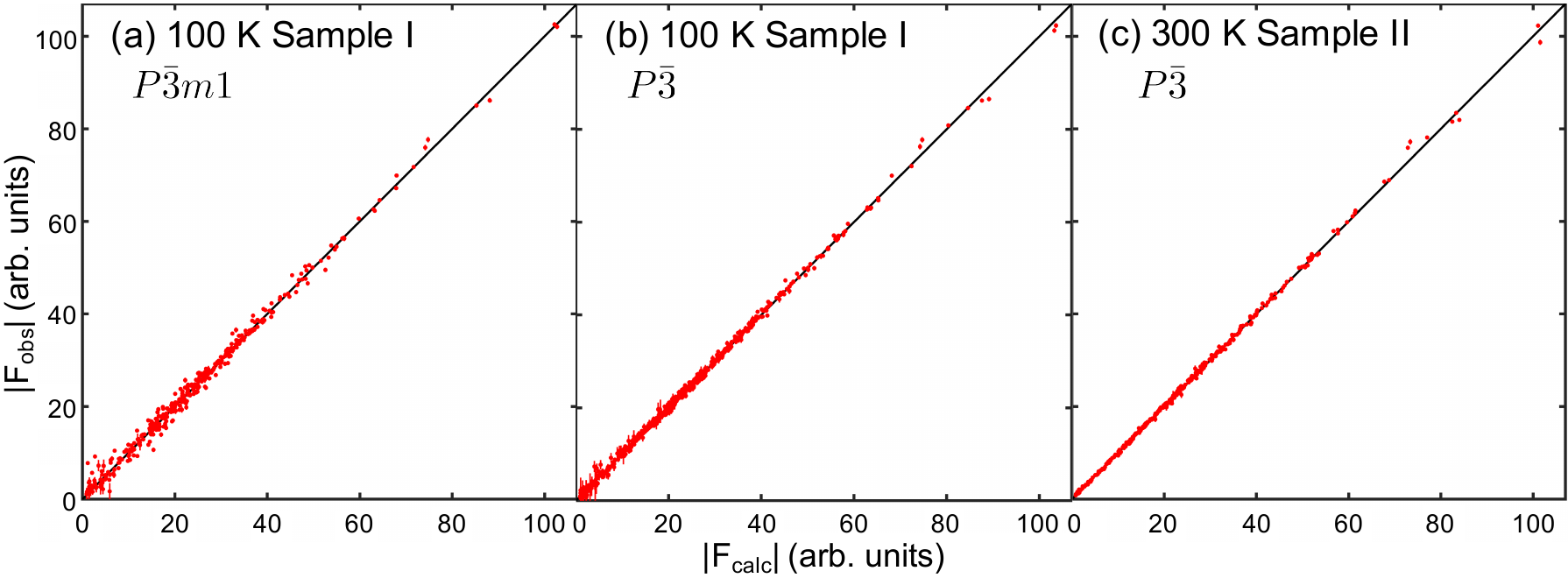}
\caption{ 
Observed versus calculated x-ray structure factor magnitudes for (a) the nominal $P\bar{3}m1$ crystal structure in Fig.~{\ref{F:ADPs}}(a) and Table~{\ref{T:XRD_100K_P-3m1}}, (b) the lower-symmetry $P\bar{3}$ structure in Fig.~{\ref{F:ADPs}}(b) and Table~{\ref{T:XRD_100K_P-3}}, and (c) the $P\bar{3}$ structure in Table~{\ref{T:XRD_P-3}}. Panels (a) and (b) correspond to the same data collected for the twinned Sample~I at 100~K, and (c) is for data collected on the almost untwinned Sample~II at room temperature. $|F_\text{obs}|$ is the experimental structure factor magnitude corrected for extinction as explained in Table~{\ref{T:XRD_100K_P-3m1}}. Error bars represent one standard deviation. Only reflections with $I>2\sigma(I)$ defined as observed peaks are included in the refinements, but all points with non-zero intensity are included in the plot. The solid line shows the 1:1 agreement in all panels.
}\label{F:1to1}
\end{figure*}

\begin{table}[tbh]
\caption{Fractional atomic coordinates and displacement parameters deduced from refinement of single crystal x-ray diffraction data at 100~K from Sample~I (size $0.056\times 0.077\times 0.183$~mm$^3$) against the nominal structural model with space group $P\bar{3}m1$ (No.~164), with the result illustrated in Fig.~{\ref{F:ADPs}}(a). Estimated standard deviations are given in parentheses. Anisotropic $U_{ij}$ displacement parameters and equivalent isotropic $U_\text{eq}$ parameters are in units of 10$^{-3}$\AA$^2$, with $U_\text{eq}$ defined as $(U_{11} + U_{22} + U_{33})/3$. Lattice parameters are $a$ = $b$ = 5.3054(1)~\AA, $c$ =6.9779(1)~\AA, $R_\mathrm{int}$=4.6\%, total number of unique reflections $N$=371 out of which $N(I>2\sigma(I))$=356 included in the refinement, $R(I>2\sigma(I))$ =3.81\%, $wR(I>2\sigma(I))$ =8.94\%, $S$ = 1.013. Extinction corrections were applied by fitting the raw experimental structure factor magnitudes to $|F_\text{calc}|/\left(1+0.001\xi|F_\text{calc}|^2\lambda^3/\sin2\theta\right)^{1/4}$, where $F_\text{calc}$ is the calculated structure factor, $\lambda=0.71073$~\AA~is the x-ray wavelength and $2\theta$ is the total scattering angle where the reflection is observed; $\xi$ was refined to 0.058(7).}
\begin{ruledtabular}
\begin{tabular}{cccccc}
Site&
Wyckoff&
$x$&
$y$&
$z$&
$U_\text{eq}$\\
\colrule
Ba & $1a$ & 0 & 0 & 0 & 4.1(2)\\
Co& $1b$ & 0 & 0 & 1/2 & 3.4(3)\\
P& $2d$ & 1/3 & 2/3 & 0.2414(3) & 3.4(4)\\
Na& $2d$ & 1/3 & 2/3 & 0.6793(6) & 8.0(7)\\
O2 & $2d$ & 1/3 & 2/3 & 0.0237(9) & 6.3(1.1)\\
O1 & $6i$ & 0.1792(6) & -0.1792(6) &  0.3189(6)  & 49(3)\\
\colrule
$U_{11}$&$U_{22}$&$U_{33}$&$U_{23}$&$U_{13}$&$U_{12}$\\
\colrule
5.2(3) & 5.2(3) & 2.0(3) & 0 & 0 & 2.6(1)\\
4.0(4) & 4.0(4) & 2.1(6) & 0 & 0 & 2.0(2) \\
4.0(5) & 4.0(5) & 2.1(7) & 0 & 0 & 2.0(3) \\
8.3(1.0) & 8.3(1.0) & 7.2(1.6) & 0 & 0 & 4.2(5)\\
8.2(1.6) & 8.2(1.6) & 2(2) & 0 & 0 & 4.1(8)\\
107(6) & 107(6) & 4.4(1.6) & -0.5(7) & 0.5(7) & 106(6)\\
\end{tabular}
\end{ruledtabular}
\label{T:XRD_100K_P-3m1}
\end{table}

\begin{table}[tbh]
\caption{
Same as Table~{\ref{T:XRD_100K_P-3m1}}, but for refinement of the 100~K data from Sample~I against the reduced-symmetry space group $P\bar{3}$ (No.~147), with the result illustrated in Fig.~{\ref{F:ADPs}}(b). Here the oxygen O1 is located at the general position, whereas it occupied a special position $6i (x,\bar{x},z)$ in the higher-symmetry structure listed in Table~{\ref{T:XRD_100K_P-3m1}}. Refinement parameters $R_\mathrm{int}$ =4.4\%, $N$=589 reflections, $N(I>2\sigma(I))$=564, $R(I>2\sigma(I))$ =1.77\%, $wR(I>2\sigma(I))$ =3.96\%, $S$=1.027, $\xi$=0.046(3). 
The refined population of the minority twin, related by 2-fold rotation around the $\bm{a}+\bm{b}$ axis, was 46.5(4)\%. }
\begin{ruledtabular}
\begin{tabular}{cccccc}
Site&
Wyckoff&
$x$&
$y$&
$z$&
$U_\text{eq}$\\
\colrule
Ba & $1a$ & 0 & 0 & 0 & 4.03(8)\\
Co& $1b$ & 0 & 0 & 1/2 & 2.67(1)\\
P& $2d$ & 1/3 & 2/3 & 0.24152(1) & 2.7(1)\\
Na& $2d$ & 1/3 & 2/3 & 0.6788(2) & 6.8(2)\\
O2 & $2d$ & 1/3 & 2/3 & 0.0234(3) & 6.1(4)\\
O1 & $6g$ & 0.2347(3) & -0.1236(3) &  0.318(2)  & 5.7(3)\\
\colrule
$U_{11}$&$U_{22}$&$U_{33}$&$U_{23}$&$U_{13}$&$U_{12}$\\
\colrule
5.18(9) & 5.18(9) & 1.7(1) & 0 & 0 & 2.59(5)\\
3.07(14) & 3.07(14) & 1.9(2) & 0 & 0 & 1.54(7)\\
2.88(19) & 2.88(19) & 2.2(3) & 0 & 0 & 1.44(9)\\
6.6(3) & 6.6(3) & 7.3(5) & 0 & 0 & 3.28(17)\\
7.4(5) & 7.4(5) & 3.5(8) & 0 & 0 & 3.7(3)\\
7.1(7) & 7.2(7) & 5.0(6) & 0.0(5) & 1.5(4) & 5.3(7)\\
\end{tabular}
\end{ruledtabular}
\label{T:XRD_100K_P-3}
\end{table}

\begin{table}[tbh]
\caption{
Same as Table~{\ref{T:XRD_100K_P-3}}, but for refinement of room temperature data from Sample~II (size $0.15\times0.17\times0.27$~mm$^3$) against the $P\bar{3}$ structural model. Lattice parameters are $a = b = 5.3162(2)$~\AA, $c = 7.0052(3)$~\AA. Refinement parameters $R_\mathrm{int}$=4.32\%, $N$=311 reflections, $N(I>2\sigma(I))$=309, $R(I>2\sigma(I))$=1.23\%, $wR(I>2\sigma(I))$=3.07\%, $S$=1.066, $\xi$=0.59(2), minority twin population 8.4(3)\%.}
\begin{ruledtabular}
\begin{tabular}{cccccc}
Site&
Wyckoff&
$x$&
$y$&
$z$&
$U_\text{eq}$\\
\colrule
Ba & $1a$ & 0 & 0 & 0 & 9.9(1)\\
Co& $1b$ & 0 & 0 & 1/2 & 6.46(1)\\
P& $2d$ & 1/3 & 2/3 & 0.24225(9) & 6.1(2)\\
Na& $2d$ & 1/3 & 2/3 & 0.6792(2) & 17.3(3))\\
O2 & $2d$ & 1/3 & 2/3 & 0.0251(3) & 11.9(3))\\
O1 & $6g$ & 0.2312(3) & -0.1272(3) & 0.3191(2) & 13.0(4)\\
\colrule
$U_{11}$&$U_{22}$&$U_{33}$&$U_{23}$&$U_{13}$&$U_{12}$\\
\colrule
12.2(2) & 12.2(2) & 5.4(2) & 0 & 0 & 5.46(9)\\
7.0(2) & 7.0(2) & 5.3(3) & 0 & 0 & 2.86(10)\\
6.6(2) & 6.6(2) & 5.1(3) & 0 & 0 & 2.59(13)\\
16.5(4) & 16.5(4) & 18.9(7) & 0 & 0 & 7.5(2)\\
16.8(6) & 16.8(6) & 5.2(9) & 0 & 0 & 8.4(3)\\
14.4(6) & 13.4(6) & 11.5(5) & -0.4(4) & 3.3(4) & 9.6(5)\\
\end{tabular}
\end{ruledtabular}
\label{T:XRD_P-3}
\end{table}

The above analysis suggests either an intrinsic structural disorder at the O1 oxygens coordinating the cobalt ions, or a lower-symmetry crystal structure. To distinguish between those two scenarios, we have performed refinements of the same diffraction data against a reduced-symmetry $P\bar{3}$ structural model that allows for a rotation of the CoO$_6$ octahedra by a variable angle $\psi$ around the $z$-axis passing through the Co site. We assumed co-existing $P\bar{3}$ twins with octahedral rotation angles $\pm \psi$, or equivalently, twins related by a mirror normal to any of the $\bm{a}$, $\bm{b}$ or $\bm{a}+\bm{b}$ axes, or a 2-fold rotation around any of those axes. This refinement resulted in much smaller oxygen ADPs, as illustrated in Fig.~\ref{F:ADPs}(b) and listed in Table~{\ref{T:XRD_100K_P-3}}, with a significant improvement in the agreement [compare Figs.~\ref{F:1to1}(b) and (a)]. To further test the symmetry reduction scenario, several crystals extracted from the same growth batch were systematically screened at room temperature in an attempt to find a predominantly single-grain sample. Fig.~\ref{F:1to1}(c) shows refinement results for one such sample, Sample~II; the quantitative refinement agreement is excellent, whereas a fit to the nominal $P\bar{3}m1$ model is significantly worse [not shown, $R_{\rm{int}}$ and $wR(I>2\sigma(I))$ increase from 4.32 and 3.07\%, to 11.6 and  6.42\%, respectively]. In the refined $P\bar{3}$ structural model, the CoO$_6$ octahedra are rotated around the $c$-axis by an angle $\psi=10.15(4)^\circ$ at 100~K.\\ 

For the multi-crystal sample mount used in the INS experiments illustrated in Fig.~\ref{F:samplemount}, the axis orientations of each crystal piece were determined using single crystal x-ray diffraction, collecting patterns by directing the x-ray beam to scatter from a point on the edge, with several such points tested for consistency. The orientation of axes was then deduced from comparison with the expected diffraction pattern for the nominal $P\bar{3}m1$ space group, as that provided a good approximation of the diffraction pattern for samples with co-existing $P\bar{3}$ twins related by a two-fold rotation around the $\bm{a}$-axis, which tended to be the case for the large crystals.

\section{Fits to a Heisenberg-Kitaev Hamiltonian}\label{A:JKfits}
\begin{figure}[b]
\includegraphics[width=0.48\textwidth]{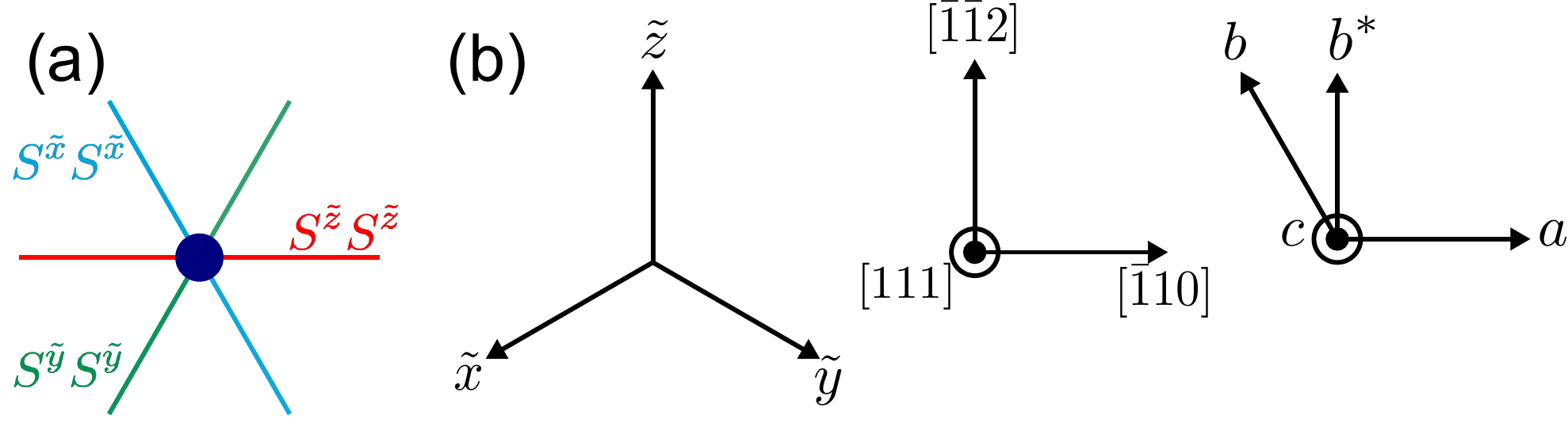}
\caption[Definition of the Kitaev axes on a triangular lattice]{Definition of the Kitaev axes for the $\mathcal{H}_{JK}$ Hamiltonian in Eq.~(\ref{E:hamJK}). (a) Nearest-neighbor bonds are color coded blue, green, red according to the corresponding Kitaev axis $\tilde{x}$, $\tilde{y}$ or $\tilde{z}$. (b) Projection of the cubic $\tilde{x}$, $\tilde{y}$ and $\tilde{z}$ axes onto the crystallographic $ab$ plane and relation between the orthogonal crystallographic directions $a$, $b^*$ and $c$ and the cubic axes.}\label{F:triangularkitaev}
\end{figure}

Here we discuss comparison of the high-field dispersion data to an alternative spin Hamiltonian to the XXZ model, in particular the Heisenberg-Kitaev model proposed in Ref.~\cite{Wellm2021}. This model is defined as
\begin{equation}\label{E:hamJK}
\mathcal{H}_{JK}=\sum_{\langle ij\rangle}J\mathbf{S}_i\cdot\mathbf{S}_j+KS_i^{\gamma}S_j^{\gamma},
\end{equation}
where the second term is the Kitaev exchange, with the bond-dependent Ising axes for the nearest-neighbor bonds as indicated by the bond color in Fig.~\ref{F:triangularkitaev}(a). The Ising axes $\gamma$ form a set of cubic axes $\tilde{x}\tilde{y}\tilde{z}$ illustrated in Fig.~\ref{F:triangularkitaev}(b) and defined such that 
\begin{equation}
\mathbf{a}^{\phantom{*}} \parallel  [\bar{1} 1 0], \quad \mathbf{b}^* \parallel [\bar{1} \bar{1} 2],\quad \mathbf{c} \parallel [1 1 1],\nonumber
\end{equation}
where the square brackets refer to the cubic Kitaev axes. For each bond, the Kitaev axis is perpendicular to the bond direction. 

The magnon dispersion relation in the polarized phase for $\mathcal{H}_{JK}+\mathcal{H}_{\rm{Zeeman}}$ can be obtained analytically for field along either $z$ or $y$, as discussed below, and these expressions were checked against SpinW \cite{Toth2015nu} calculations. The dispersion in field along $z$ is obtained as
\begin{align}
\hbar\omega_{\parallel}(\mathbf{Q})=&2S\sqrt{\mathcal{A}^2-|\mathcal{B}|^2}, \label{E:lswtJKc}\\
\mathcal{A} =& g_c\mu_BB_z - 3\left(J+\frac{K}{3}\right) + \left(J+\frac{K}{3}\right)\gamma\left(\mathbf{Q}\right), \nonumber\\
\mathcal{B}=&\frac{K}{2}\left[\left(\frac{1}{\sqrt{3}}-\frac{i}{3}\right)\cos{2\pi h} + \frac{2i}{3}\cos{2\pi k}\right. \nonumber \\
 & \qquad \left. + \left(-\frac{1}{\sqrt{3}}-\frac{i}{3}\right)\cos{2\pi(h+k)}\right] \nonumber,
\end{align}
with $\gamma(\mathbf{Q})$ already defined in Eq.~(\ref{E:gamma}). 

The reported INS data in Ref.~\cite{Sheng2022} is for a field $B_z$ along the $c$-axis significantly above the transition to the polarized phase, in which case $\mathcal{A}$ is the dominant term, $|\mathcal{A}|\gg|\mathcal{B}|$, and $\hbar\omega_{\parallel}(\mathbf{Q})\approx 2S\mathcal{A}$. This has the same functional form as Eq.~(\ref{E:lswtC}) with $J_z=J_{xy}=J+K/3$, i.e. as for a pure Heisenberg model with an effective exchange $J+K/3$, so $J$ and $K$ cannot be refined independently. To deduce both $J$ and $K$, one therefore needs to be away from this regime. Namely, one must perform measurements of the dispersion in polarizing field along $c$ for fields $B_z$ in very close proximity to the polarizing field and for wave vectors $\mathbf{Q}$ where $|\mathcal{B}|$ becomes comparable to $|\mathcal{A}|$, as only then is the dispersion sensitive separately to $J$ and $K$. However, if instead the field is applied transverse to the $z$-axis, the dispersion relation even much above the critical field $B_{\rm C}$ has a different functional form and is directly sensitive to the presence of a Kitaev exchange. In particular, in the polarized phase in field along $b^*$, the magnon dispersion relation is
\begin{align}
\hbar\omega_{\perp}(\mathbf{Q})&=2S\sqrt{\mathcal{A}^2-|\mathcal{B}|^2},\label{E:lswtJKy}\\
\mathcal{A} &= g_{ab}\mu_BB_y - 3\left(J+\frac{K}{3}\right)+\left(J+\frac{K}{6}\right)\cos{2\pi h}\nonumber\\ 
 & \qquad +  \left(J+\frac{5K}{12}\right)\left[\cos{2\pi(h+k)}+\cos{2\pi k}\right] \nonumber\\
\mathcal{B}&=\frac{K}{2}\left[\frac{i}{3}\cos{2\pi h} + \left(\sqrt{\frac{2}{3}}+\frac{i}{6}\right)\cos{2\pi k}\right. \nonumber \\
 &\qquad \qquad \left. +\left(\sqrt{\frac{2}{3}}+\frac{i}{6}\right)\cos{2\pi(h+k)}\right].\nonumber
\end{align}
This has a different functional form compared to the XXZ dispersion in Eq.~(\ref{E:lswtB}). Therefore, if dispersion data collected in high transverse field are also taken into account, the XXZ and $JK$ scenarios can be readily distinguished, as illustrated by the poor agreement between observed and calculated magnon energies shown in Fig.~\ref{F:onetoone}(b) for the $JK$ model compared to the excellent agreement for the XXZ model in Fig.~\ref{F:onetoone}(a). 
\begin{table}
\centering
\caption{
Best-fit Hamiltonian parameters using the Heisenberg-Kitaev exchange model in Eqs.~(\ref{E:hamJK}) and (\ref{E:Zeeman}) with the level of agreement for magnon energies illustrated in Fig.~\ref{F:onetoone}(b). 
}\label{T:hamparamsJK}
\begin{tabular}{cd}
\hline
\hline
$J$ & 0.0900(18)~\text{meV}\\
$K$ & 0.0063(14)~\text{meV}\\
$g_c$ & 4.230(10) \\
$g_{ab}$ & 4.340(17) \\
$\delta B$ & 0.142(17)~\text{T at $B=1.7~$T}\parallel b^*\\
\hline
\hline
\end{tabular}
\end{table}

The presence of a Kitaev term ($K$) or next-nearest-neighbor in-plane exchange ($J_3$) as small perturbations to the minimal XXZ Hamiltonian was also investigated. Allowing either of these terms did not, however, significantly improve the fits, and gave values of $K$ and $J_3$ at most of order 5\% of $J_z$, i.e., the same order of magnitude as the estimated magnitude of the neglected interlayer couplings as discussed in Appendix~\ref{A:NBCPO_interlayer}, and placing this material well away from the spin-liquid regime \cite{gallegos2025phasediagrameasyaxistriangularlattice}. Therefore, in the analysis of the excitation spectrum in the main text, the XXZ Hamiltonian parameters in Table~\ref{T:hamparams} are used, as these already provide an excellent quantitative description of the high field dispersions in field along two orthogonal directions, as illustrated in Fig.~\ref{F:onetoone}(a).

\section{\label{A:2M} Two-magnon excitations in LSWT}
Here we provide technical details of the calculation of the dynamical correlations for one- and two-magnon scattering processes for the non-collinear 3-sublattice magnetic structure at zero and intermediate transverse field. We are not claiming the formulas listed here to be original, but are including them here for completeness; the approach used is similar to that employed in Ref.~\cite{Sheng2024}.  
It is convenient to introduce a local reference frame $x'y'z'$, with $z'$ along the local ordered spin at each site. The transverse directions $x', y'$ are defined according to the convention used by the SpinW package \cite{Toth2015nu} relevant to the present context, i.e., $\mathbf{y'}\parallel(\mathbf{z'}\times\mathbf{a})$, and $\mathbf{x'}\parallel (\mathbf{y'}\times\mathbf{z'})$. 
In this local frame, all spins point up along the $z'$ axis. The magnetic structure of the XXZ TLAF in all transverse fields below $B_{\rm{C}}$ is a non-collinear 3-sublattice order, see Sec.~\ref{S:phasediagram}, and the transformed spin Hamiltonian has the periodicity of the enlarged 3-site unit cell of the magnetic structure. This is because the unique site out of the three in the magnetic unit cell [in any of the Y, iY, $\Psi$ or V phases in Fig.~\ref{F:phasediagram}(b) and (c)] has a different local environment compared to the other two sites. This is in contrast to the case of spiral ordered phases, such as the 120$^{\circ}$ order in the easy-plane XXZ TLAF model \cite{Macdougal2020} or the incommensurate spiral order in the distorted (isosceles) triangular lattice \cite{Coldea2003}, where each magnetic site in the local frame has the same local environment, so the transformed spin Hamiltonian has the periodicity of the structural cell and the problem is reduced to a single magnetic sublattice. In the present case no such simplification occurs, and one must consider the full 3-sublattice problem.    

Two-magnon scattering processes involve two magnon-creation operators so in the local frame, these occur entirely in the longitudinal ($z'z'$) polarization. For an isotropic $g$-factor at zero temperature, the latter is given by 
\begin{equation}
S^{z'z'}(\mathbf{Q},\omega)=g^2\mu_B^2\sum_{\lambda_f}|\langle\lambda_f|S^{z'}(\mathbf{Q})|\text{GS}\rangle|^2\delta\left(E_{\lambda_f}-\hbar\omega\right),\nonumber
\end{equation}
where the sum extends over all states $|\lambda_f\rangle$ of energy $E_{\lambda_f}$ relative to the ground state $|\text{GS}\rangle$ and where $S^{z'}(\mathbf{Q})=(1/\sqrt{\mathcal{N}})\sum_{\mathbf{R}}\exp(i\mathbf{Q}\cdot\mathbf{R})S^{z'}_{\mathbf{R}}$,  with $\mathbf{R}$ running over the positions of the $\mathcal{N}$ magnetic sites in the lattice. 
$S^{z'z'}$ has two contributions: at zero energy, the magnetic Bragg peaks, and at finite energy, two-magnon excited states. Transforming back to the original global frame, the two-magnon contribution to the $\zeta\eta$ component of the dynamical structure factor normalized per site is given by 
\begin{widetext}
\begin{align}
S^{\zeta\eta}_{2\rm{M}}(\mathbf{Q},\omega)= 
&\frac{Z_{2\rm{M}}g^{\zeta}g^{\eta}\mu_B^2}{2{N_{\mathrm{S}}}\mathcal{N}^2}\sum_{\mathbf{k},\mathbf{k'},\mathbf{\tau}}\delta\left(\mathbf{Q}-\mathbf{k}-\mathbf{k'}-\mathbf{\tau}\right)\sum_{j,j'=1}^{N_{\mathrm{S}}}\delta\left(\hbar\omega-\hbar\omega_{j}(\mathbf{k})-\hbar\omega_{j'}(\mathbf{k'})\right)\sum_{m,m'=1}^{N_{\mathrm{S}}} \mathcal{C}_{mjj'}^{\zeta}(\mathbf{k},\mathbf{k'})^* \, \mathcal{C}_{m'jj'}^{\eta}(\mathbf{k},\mathbf{k'}),\label{E:Sab}\\
\mathcal{C}_{mjj'}^{\zeta}(\mathbf{k},\mathbf{k'})=&f^{m\phantom{'}}_{z'\zeta}e^{2\pi i(\mathbf{k}+\mathbf{k'})\cdot \mathbf{r}_m}\left[u_{mj}(\mathbf{k})v_{mj'}(\mathbf{k'})+u_{mj'}(\mathbf{k'})v_{mj}(\mathbf{k})\right],\nonumber
\end{align}
\end{widetext}
where $\zeta,\eta = x, y$ and $z$, and where we have allowed for an anisotropic $g$-tensor assumed to be diagonal in the $xyz$ axes frame, which is the case for the experiments reported here. In the above equations, $\mathbf{k}$ and $\mathbf{k'}$ are the wave vectors of the two individual magnons created in a two-magnon scattering process, with each running over the full structural Brillouin zone, $j,j'$ label the different magnon modes, $m,m'$ run over the distinct magnetic sublattices in the magnetic unit cell, $\hbar\omega_j(\mathbf{k})$ is the energy of the $j$th magnon mode with wave vector $\mathbf{k}$, $f^{m}_{z'\zeta}$ is the component of the $\hat{z}'$ axis of the $m$th sublattice along the global $\zeta$ direction, $\mathbf{r}_m$ is the position of the $m$th sublattice in the reference magnetic unit cell and $N_{\mathrm{S}}$ is the number of sublattices. 
The first $\delta$-function in Eq.~(\ref{E:Sab}) indicates that the wave vector $\mathbf{Q}$ is equivalent to the sum of the two wave-vector transfers $\mathbf{k}$ and $\mathbf{k'}$ up to a reciprocal lattice vector $\mathbf{\tau}$. The overall pre-factor $Z_{2\rm{M}}$ is introduced here to allow for an overall effective renormalization of the intensity due to quantum fluctuations beyond LSWT, for which $Z_{2\rm{M}}=1$.
The quantities $u_{mj}(\mathbf{k})$, $v_{mj}(\mathbf{k})$ transform between the boson operators $a_{\mathbf{k},m}$ for the $N_{\mathrm{S}}$ different magnetic sublattices, $m=1$ to $N_{\mathrm{S}}$, and the normal mode boson operators $b_{\mathbf{k},j}$ as
\begin{align}
a_{\mathbf{k},m} = \sum_{j=1}^{N_{\mathrm{S}}} u^{*}_{mj}(-\mathbf{k})b_{\mathbf{k},j} + v_{mj}(\mathbf{k})b_{-\mathbf{k},j}^{\dagger},\nonumber\\
a_{-\mathbf{k},m}^{\dagger} = \sum_{j=1}^{N_{\mathrm{S}}} v^*_{mj}(-\mathbf{k})b_{\mathbf{k},j} + u_{mj}(\mathbf{k})b_{-\mathbf{k},j}^{\dagger},\nonumber
\end{align}
where $j$ runs over all the magnon modes, with ${N_{\mathrm{S}}}$=3 for $B< B_{\rm{C}}$ and ${N_{\mathrm{S}}}=1$ for $B\geq B_{\rm{C}}$, see Fig.~\ref{F:phasediagram}(c). 

One-magnon processes occur entirely in the transverse dynamical correlations $S^{x'x'}$, $S^{y'y'}$, $S^{x'y'}$ and $S^{y'x'}$ as they are due to matrix elements of single magnon-creation operators. The one-magnon contribution to the dynamical structure factor normalized per site is therefore
\begin{widetext}
 \begin{align}
S^{\zeta\eta}_{1\rm{M}}(\mathbf{Q},\omega)=& \frac{Z_{1\rm{M}}Sg^{\zeta}g^{\eta}\mu_B^2}{2{N_\mathrm{S}}^2\mathcal{N}}\sum_{\mathbf{k},\mathbf{\tau}}\delta\left(\mathbf{Q}-\mathbf{k}-\mathbf{\tau}\right)\sum^{N_{\mathrm{S}}}_{j}\delta\left(\hbar\omega-\hbar\omega_j(\mathbf{k})\right) \sum^{N_{\mathrm{S}}}_{m,m'}\tilde{\mathcal{C}}^{\zeta}_{mj}(\mathbf{k})^* \, \tilde{\mathcal{C}}^{\eta}_{m'j}(\mathbf{k}),\label{E:Sab_1m}\\
\tilde{\mathcal{C}}^{\zeta}_{mj}(\mathbf{k})=&e^{2\pi i(\mathbf{k}+\mathbf{k'})\cdot \mathbf{r}_m}\left(f^{m\phantom{'}}_{x'\zeta} \left[u_{mj}(\mathbf{k})+v_{mj}(\mathbf{k})\right] +  f^m_{y'\zeta} i\left[v_{mj}(\mathbf{k})-u_{mj}(\mathbf{k})\right]\right),\nonumber
\end{align}
\end{widetext}
where $Z_{1\rm{M}}$ is an overall one-magnon intensity renormalization factor due to quantum fluctuations beyond LSWT, for which $Z_{1\rm{M}}=1$, and $f^{m}_{\{x',y'\}\zeta}$ is the component of $\hat{x}',\hat{y}'$ on the $m$th sublattice along the global $\zeta$ direction.

\begin{table*}[tb]
\centering
\caption{The sum total of the intensity contribution in each channel in the local $x'y'z'$ frame in units of $(g\mu_B)^2$ assuming an isotropic $g$-tensor, as calculated in LSWT ($Z_{1\rm{M}}=Z_{2\rm{M}}=1$), at some representative fields. The first and second rows correspond to the two different polarizations of single magnon scattering. The third row shows the longitudinal polarization $M^{z'z'}$, which has contributions from magnetic Bragg peaks (BP) and from two-magnon (2M) excitations, and which should sum to 1/4 as per Eq.~(\ref{E:sumrule}). The last two rows show the two contributions separately, with the fourth row listing the Bragg peak contribution as per Eq.~(\ref{E:BPlswt}) and the fifth row giving the two-magnon contribution, calculated using Eq.~(\ref{E:Sab}). At low field, the sum rules are only very approximately satisfied, but the relative intensities of the single-magnon and two-magnon scattering are nevertheless close to what is expected from the sum rules. The 3.5~T column shows that well into the paramagnetic phase, linear spinwave theory is an asymptotically good approximation as the sum rules are very close to being exactly satisfied. Any departure is attributed to the effects of quantum fluctuations due to spin-non-conservation because the field is applied perpendicular to the Ising axis. At each field, $Z_{\text{1M}}$ was chosen such as to satisfy the moment sum rule in Eq.~(\ref{E:xysumrule}) and $Z_{2\rm{M}}$ was chosen to satisfy the sum rule for $M^{z'z'}$ assuming that the Bragg peak contribution in Eq.~(\ref{E:BPlswt}) is un-renormalized.
}
\begin{tabular}{ccccccccc}
\hline
\hline
\vspace{0.5mm}
                       & 0~T  & 0.25~T  & 0.7~T & 1~T & 1.4~T & 1.7~T & 3.5~T & Sum rules\\
$M^{x'x'}$            & 0.3450 & 0.3045 & 0.2756& 0.2630&0.2477 & 0.2253&0.2482 & 0.25 \\
$M^{y'y'}$            & 0.3208 & 0.2650 & 0.2531& 0.2532&0.2641 & 0.2968&0.2522 & 0.25\\
$M^{z'z'}$            & 0.3141 & 0.2624 & 0.2528& 0.2510&0.2511 & 0.2561&0.2499 & 0.25\\
$M^{z'z'}_{\text{BP}}$& 0.1153 & 0.1855 & 0.2222& 0.2342&0.2384 & 0.2284&0.2496 \\
$M^{z'z'}_{2\rm{M}}$  & 0.1988 & 0.0769 & 0.0306& 0.0169&0.0127 & 0.0277&0.0003 \\
\hline
\hline
\end{tabular}
\label{T:normalization}
\end{table*}

The integral equations for the dynamical correlations $S^{\zeta\eta}(\mathbf{Q},\omega)$ in Eq.~(\ref{E:Sab}) were calculated numerically using a Monte-Carlo method as in Refs.~\cite{Tennant95,Coldea2003}, using $10^8$ $(\mathbf{k},\mathbf{k}^{\prime})$ wave-vector pairs, where each wave vector was selected randomly inside the Brillouin zone. At each given magnetic field, the SpinW package \cite{Toth2015nu} was used to calculate the local spin projection components $f^{m}_{z'\zeta}$ and the functions $u_{mj}(\mathbf{k})$, $v_{mj}(\mathbf{k})$, $\hbar\omega_j(\mathbf{k})$. The mean-field magnetic structure was either found iteratively using SpinW, or through analytic calculations. 
The intensity prefactors $Z_{1\mathrm{M}}$ and $Z_{2\mathrm{M}}$ were determined such that the dynamical correlations satisfied the total moment sum rules in the local frame.
These rules state that for a spin-half system with an isotropic $g$-tensor \cite{Lorenzana2005},
\begin{equation}\label{E:sumrule}
M^{\zeta\eta}=\frac{1}{\mathcal{N}}\sum_{\mathbf{Q}} \int \mathrm{d}(\hbar\omega) S^{\zeta\eta}\left(\mathbf{Q},\omega\right)= \frac{1}{4}(g\mu_B)^2\delta_{\zeta\eta},
\end{equation}
where the sum is over all wave vectors $\mathbf{Q}$ in the full structural Brillouin zone. This result is independent of the spin exchange Hamiltonian, magnetic field or temperature. The contributions to $M^{x'x'}$ and $M^{y'y'}$ come from single-magnon excitations. In the fixed frame, it is not possible to normalize these polarizations separately, so a common overall scale factor $Z_{1\mathrm{M}}$ has been assumed for all 1-magnon contributions, such that
\begin{equation}\label{E:xysumrule}
M^{x'x'}+M^{y'y'}=\frac{1}{2}(g\mu_B)^2.  
\end{equation}
There are two contributions to $M^{z'z'}$: the magnetic Bragg peaks on the elastic line, and the inelastic part due to two-magnon scattering, as described above. For the case of three sublattices relevant here, we obtain that the magnetic Bragg peaks have total intensity per magnetic site 
\begin{equation}\label{E:BPlswt}
M^{z'z'}_{\text{BP}} = \left(S^2 - 2S\overline{\Delta S} + \overline{(\Delta S)^2}\right)(g\mu_B)^2,
\end{equation}
where
\begin{equation}
(\Delta S)_m = \langle S-S^{z'} \rangle_m = \frac{1}{\mathcal{N}N_{\mathrm{S}}} \sum_{\mathbf{k}}\sum_j |v_{mj}|^2\nonumber
\end{equation}
is the spin reduction on sublattice $m$, and where the overlines denote averages across the different magnetic sublattices. Setting $S=1/2$, the two magnon contribution to the scattering is therefore required to sum to $\left(\overline{\Delta S} - \overline{(\Delta S)^2}\right)(g\mu_{\rm B})^2$. It was found that without renormalization, i.e., assuming $Z_{1\rm{M}}=Z_{2\rm{M}}=1$, the sum rules were overestimated for all three polarizations in zero field by about 30\%, as shown in Table~\ref{T:normalization}. 

We find that, for the Hamiltonian parameters in Table~\ref{T:hamparams}, the ground state order is strongly non-collinear for all studied fields below $B_{\rm C}$, with the consequence that the two-magnon scattering has finite contributions from all six possible types of magnon pairs $\alpha\alpha$, $\beta\beta$, $\gamma\gamma$, $\alpha\beta$, $\beta\gamma$, $\alpha\gamma$, but with the relative contributions being field dependent. Here $\alpha$, $\beta$, $\gamma$ are the three magnon modes in order of increasing energy. For example, in zero field, the largest two-magnon weight (largest fraction of $M^{z'z'}_{\rm 2M}$) comes from $\alpha\alpha$ and $\beta\beta$ continua, whereas at 1.4~T the dominant contribution comes from the $\alpha\alpha$ continuum. 

\begin{figure}[htb]
    \includegraphics[width=0.4\textwidth]{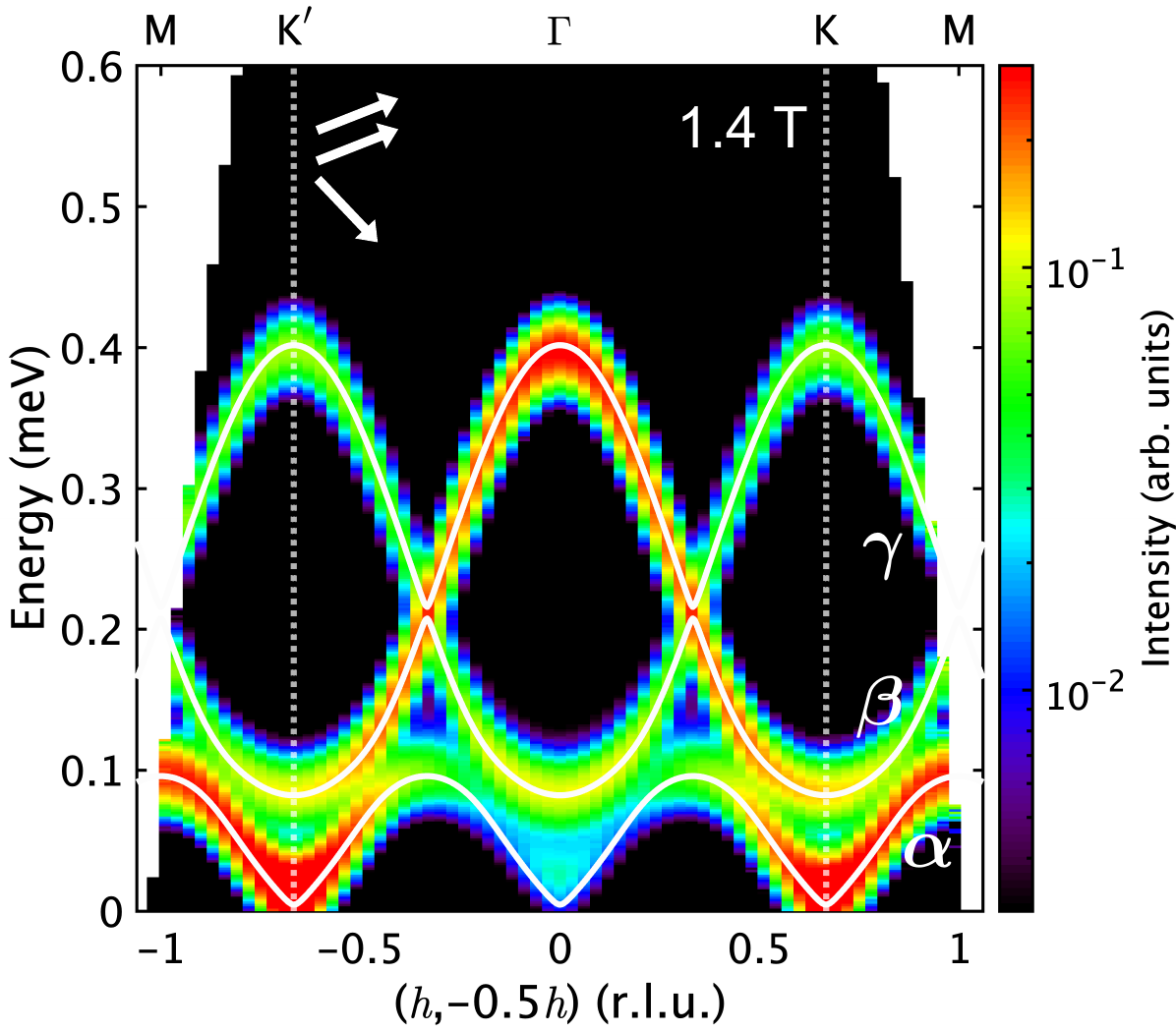}
    \caption{Calculated INS spectrum, including both one- and two-magnon excitations, for the model including a small biquadratic exchange to stabilize a V-phase at 1.4~T, as described in the text, to be compared with calculations for a $\Psi$ phase in Fig.~\ref{F:twomagnon}(h) with all symbols having the same meaning and intensities plotted on a log scale.}
    \label{F:Vphase}
\end{figure}

The two-magnon contribution is included in all the calculation plots in Fig.~\ref{F:twomagnon} (even columns), as well as in Figs.~\ref{F:degeneracy}(b, c) and \ref{F:Vphase}. In these calculations, the plotted intensity is the total dynamical structure factor $\mathcal{S}=\sum_{\zeta\eta}(\delta_{\zeta\eta}-\hat{Q}_\zeta\hat{Q}_\eta)S^{\zeta\eta}$, where $\hat{Q}_{\zeta}=Q_\zeta/|\mathbf{Q}|$ and $\mathbf{Q}$ is the total wave-vector transfer, taking into account the different $\mathbf{Q}$ values at all contributing pixels. The two-magnon signal is most prominent at zero field and decreases in magnitude upon increasing field, but is still present at all probed fields. For example, in Fig.~\ref{F:twomagnon}(b) (zero field), the blue shaded signal above the $\gamma$ mode near K and K$^\prime$ points is a two-magnon continuum of mainly $\beta\gamma$ character, whereas the green shaded region between $\beta$ and $\gamma$ modes at the same momenta is a continuum of mainly $\alpha\alpha$ character. In Fig.~\ref{F:twomagnon}(f) at 0.25~T, the blue shaded continuum between 0.25 to 0.35~meV comes mainly from a $\beta\beta$ continuum, and in Fig.~\ref{F:twomagnon}(j) at 0.7~T, the blue regions near K and K$^\prime$ around 0.3 and 0.45~meV come from $\beta\beta$ and $\gamma\gamma$ continua, respectively. Furthermore, the $\beta$ and $\gamma$ modes at 0.7 T overlap with $\alpha\alpha$ and $\alpha\beta$ continua, not visible on the plotted scale. In Figs.~\ref{F:twomagnon}(d) (1 T) and (h) (1.4 T), the continuum is only just visible: in (d) the purple region around 0.3 meV around the K and K$^{\prime}$ points is a continuum of $\beta\beta$ character and in (h) the blue region around 0.05 meV near the $\Gamma$ point is an $\alpha\alpha$ continuum. The $\beta$ and $\gamma$ modes at 1 and 1.4 T also overlap with $\alpha\alpha$, $\alpha\beta$, and $\beta\beta$ continua, which are not visible on the plotted scale. At the end of Sec.~\ref{S:lowINSexpcalc}, we analyze in turn where magnon modes overlap with continua at all experimentally measured fields and list kinematically-allowed decay channels. 

\section{Spectrum in the V- vs. $\Psi$-phase below $B_{\rm C}$}\label{A:Vphase}
Theoretical calculations that take into account quantum fluctuations predict that the spontaneous magnetic order below $B_{\rm C}$ is a V phase, rather than the $\Psi$ phase predicted semi-classically, with the relevant phase diagrams compared in Figs.~\ref{F:phasediagram}(b) and (c). In this Appendix, we present an approximate calculation of the spectrum in the V phase in the region just below $B_{\rm C}$ and compare with the spectrum of the $\Psi$ phase at the same field. In this regime, we find that it is possible to stabilize the V-phase at the mean-field level by adding a biquadratic exchange term 
$\mathcal{H}_{\rm bq}=-J_{\rm{bq}}\sum_{\langle ij \rangle}\left(\mathbf{S}_i\cdot\mathbf{S}_j\right)^2$ with $J_{\rm{bq}}>0$, 
as an empirical way to approximately parameterize the effects of quantum fluctuations, as explained in Refs.~\cite{Chubukov1991,Griset2011}. 
We derived analytic expressions for the ground state energy for the total Hamiltonian $\mathcal{H}_{\text{total}}=\mathcal{H}_{\rm XXZ}+\mathcal{H}_{\rm Zeeman}+\mathcal{H}_{\rm bq}$ for both a V-phase (parameterized by two angles) and a $\Psi$-phase (parameterized by one angle), then found numerically the angles that gave the minimum energies in both cases. By comparing the resulting energies and testing that the spinwave Hamiltonians were positive definite using SpinW~\cite{Toth2015nu}, we deduced that a transition occurs from the $\Psi$ to the V-phase when $J_{\rm bq}$ is above a threshold value. For the purpose of the calculations, we choose a value slightly larger than the threshold.  

The calculated INS spinwave spectrum for the V-phase at 1.4~T stabilized by $J_{\rm bq}=0.011$~meV is shown in Fig.~\ref{F:Vphase}. In this case, the energetic magnitude of the biquadratic exchange, i.e., the expectation value $\langle\mathcal{H}_{\rm bq}\rangle$ in the ground state, relative to the total ground state energy $\langle\mathcal{H}_{\text{total}}\rangle$ and to the reference mean-field energy of the $\Psi$-phase at $J_{\rm bq}=0$, is 0.76\% 
and 0.70\%, respectively, so the added biquadratic exchange has only a relatively small effect, but sufficient to stabilize the V-phase. The spectrum in this case is very similar in all aspects to the one calculated for the $\Psi$ phase for $J_{\rm bq}=0$ in Fig.~\ref{F:twomagnon}(h), suggesting that the key features of the spectrum at this field 
are not sensitive to whether the ground state order is a V or a $\Psi$ phase. 

We note that comparing the experimental and theoretical phase diagrams in Fig.~\ref{F:phasediagram}(a) with (b) and (c), a V phase is also expected at two other fields where the INS spectrum was measured, namely 0.7 and 1~T. We have found that at 1~T, where the required magnitude of the biquadratic exchange is 0.018~meV, the spectrum is again very similar to that of the $\Psi$ phase at the same field in the absence of $J_{\mathrm{bq}}$ (not shown). However, we were not able to use this method of including an empirical biquadratic exchange to study the excitations of the V-phase at 0.7~T, as the magnitude of the required biquadratic exchange (0.030~meV) is too large and significantly affects the physics: the $\beta$ and $\gamma$ modes become very flat (not shown), which is qualitatively different to what is seen experimentally [Fig.~\ref{F:twomagnon}(i)].

\section{\label{A:NBCPO_interlayer} Effects of interlayer couplings in the polarized phase}

Here we provide calculations of the dispersion relations and dynamical structure factors in the field-polarized phase for the XXZ exchange model for \ch{Na2BaCo(PO4)2} extended to include the effects of interlayer couplings. We then use these expressions to analyze the spectrum along the interlayer direction and thus place an upper bound on the interlayer coupling strength.

\subsection{Calculation of dispersions and dynamical correlations in linear spinwave theory}
For simplicity, we consider only the single interlayer exchange $J_b$ in Fig.~\ref{F:interlayer_interactions} as this interlayer coupling alone gives a zero-field magnetic propagation vector in the interlayer direction $q_z=1/6$, which is within the range of values reported experimentally. The spin Hamiltonian we consider is 
\begin{widetext}
\begin{align}
\mathcal{H} = &\frac{1}{2}\sum_{\mathbf{R}}\left[\sum_{\mathbf{\delta}_1} \left( J_{xy} \left(S^x_{\mathbf{R}} S^x_{\mathbf{R}+\mathbf{\delta}_1} + S^y_{\mathbf{R}}S^y_{\mathbf{R}+\mathbf{\delta}_1}\right) + J_z S^z_{\mathbf{R}}S^z_{\mathbf{R}+\mathbf{\delta}_1}\right) + \sum_{\mathbf{\delta}_2}J_b \mathbf{S}_{\mathbf{R}}\cdot\mathbf{S}_{\mathbf{R}+\mathbf{\delta}_2}
\right] -\mu_B \sum_{\mathbf{R}} g_{ab}B_{y}S^y_{\mathbf{R}},\nonumber
\end{align}
\end{widetext}
where $\mathbf{R}$ runs over the positions of magnetic sites, located at the vertices of a stacked triangular lattice, and $\mathbf{\delta}_1$ runs over the vectors connecting sites to their six nearest neighbors in the same triangular layer, i.e., $\mathbf{\delta}_1=\mathbf{a},\mathbf{b},\mathbf{a}+\mathbf{b},-\mathbf{a},-\mathbf{b},-\mathbf{a}-\mathbf{b}$. The vector $\mathbf{\delta}_2$ connects sites to neighbors in the adjacent layers above and below, as indicated in Fig.~\ref{F:interlayer_interactions}, and runs over $\mathbf{a}-\mathbf{c},\mathbf{b}-\mathbf{c},-\mathbf{a}-\mathbf{b}-\mathbf{c},-\mathbf{a}+\mathbf{c},-\mathbf{b}+\mathbf{c},\mathbf{a}+\mathbf{b}+\mathbf{c}$ in the $P\bar{3}$ crystal structure. 

\begin{figure}[htb]
\includegraphics[width=0.5\textwidth]{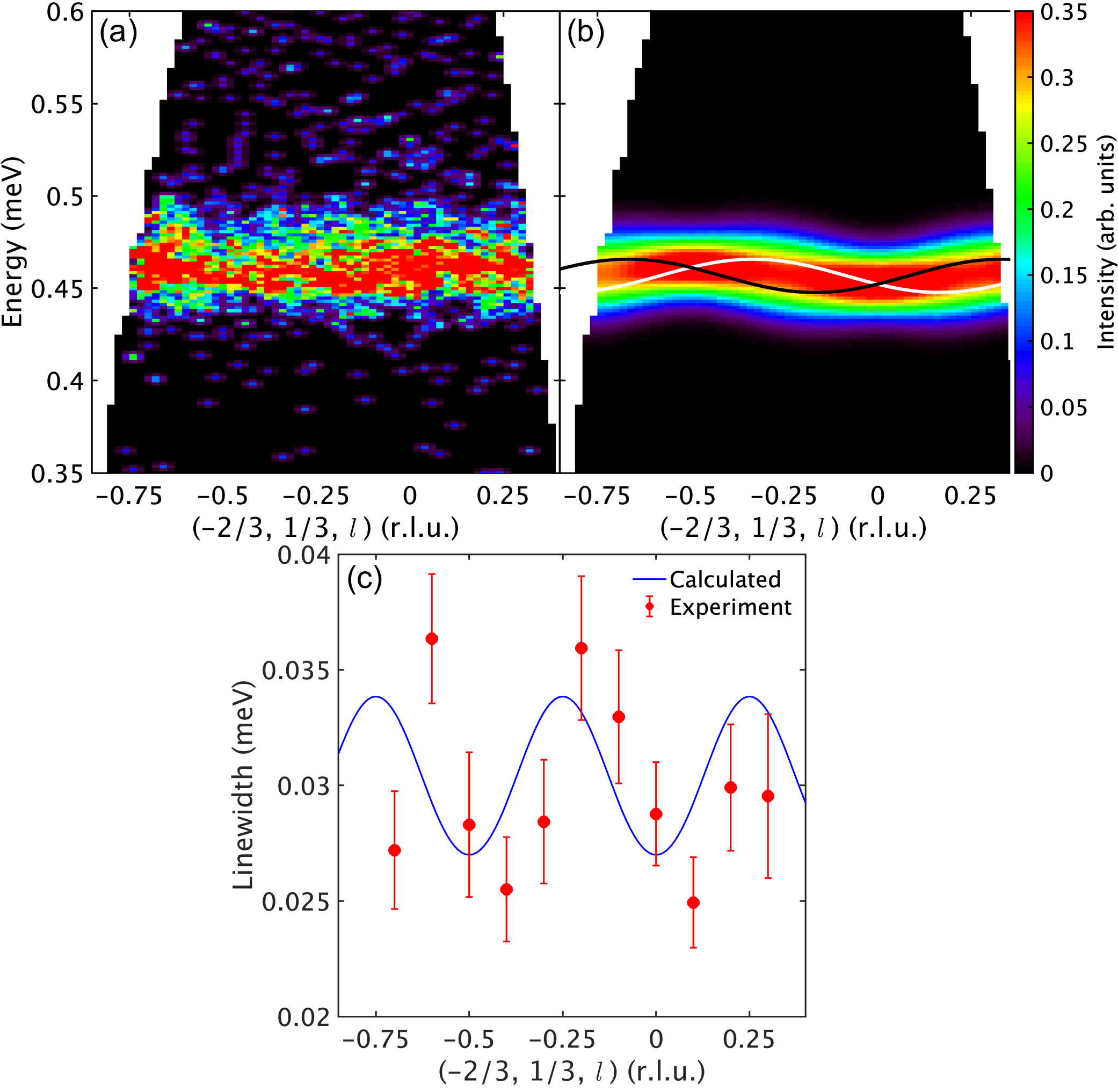}
\caption[Interlayer dispersion effects in \ch{Na2BaCo(PO4)2}]{Interlayer dispersion effects. (a) INS data taken at 3.5~T with $E_i=0.87$~meV, showing the spectrum along the $(-2/3,1/3,l)$ direction. The data have been averaged over a width of $\pm 0.025$ r.l.u. along both the $(1,-1/2,0)$ and $(010)$ directions. (b) Calculation in linear spinwave theory of the expected inelastic neutron scattering intensity, for the range of data shown in (a), taking into account the two different structural $P\bar{3}$ twins, related by the $m$ mirror, as well as the \ch{Co^{2+}} magnetic form factor. The calculation assumes $J_b=0.003$~meV and $J_2=J_a=0$. The dynamical correlations are convolved with a Gaussian of 0.027~meV FWHM to mimic instrumental resolution effects. The white and black lines are the dispersions for the main grain and the twin, respectively. In (a) and (b), color indicates INS intensity on an arbitrary scale. (c) Experimental (filled points) and calculated (solid line) linewidths (FWHM) along the interlayer direction. The experimental data points are extracted from fitting Gaussians to constant-wave-vector scans through the data in (a). The calculation uses the same parameters as in (b). The expressions for the dispersions and intensities in linear spinwave theory used in (b) and (c) are given in Eqs.~(\ref{E:lswtinterlayer}) to (\ref{E:Szzinterlayer}). 
}\label{F:interlayer}
\end{figure}

In the field-polarized phase for $B$ along $y$, the magnon dispersion in linear spinwave theory is obtained as
\begin{align}
\hbar\omega_{\perp}(\mathbf{Q})=&2S\sqrt{\mathcal{A}^2-\mathcal{B}^2},\label{E:lswtinterlayer}\\
\mathcal{A}=&g_{ab}\mu_B B_{y} - 3J_{xy} + \frac{J_z+J_{xy}}{2}\gamma(\mathbf{Q}) \nonumber \\
& \qquad +J_{b}\left[-3+\gamma_b(\mathbf{Q})\right],\nonumber\\
\mathcal{B}=&\frac{J_z-J_{xy}}{2}\gamma(\mathbf{Q}),\nonumber
\end{align}
where
\begin{equation}
\gamma_b(\mathbf{Q})= \cos{2\pi(h-l)}+\cos{2\pi(k-l)}+\cos{2\pi(-h-k-l)},\label{E:gamma_a}
\end{equation}
and $\gamma(\mathbf{Q})$ is given in Eq.~(\ref{E:gamma}). The non-zero components of the dynamical structure factor are then obtained as
\begin{align}
S^{xx}(\mathbf{Q},\omega) =g^2_{ab}\mu^2_B& \frac{\mathcal{A}+\mathcal{B}}{\hbar\omega(\mathbf{Q})}\delta(\hbar\omega-\hbar\omega_{\perp}(\mathbf{Q})),\\
S^{zz}(\mathbf{Q},\omega) =g^2_{c}\mu^2_B& \frac{\mathcal{A}-\mathcal{B}}{\hbar\omega(\mathbf{Q})}\delta(\hbar\omega-\hbar\omega_{\perp}(\mathbf{Q})),\label{E:Szzinterlayer}
\end{align}
where the $\mathbf{Q}$-dependence is implicitly assumed for $\mathcal{A}$ and $\mathcal{B}$.

\subsection{Parameterization of the interlayer dispersion}
In order to inspect the effects of the interlayer couplings, the spectrum of magnons in the direction perpendicular to the layers is plotted in Fig.~\ref{F:interlayer}(a) along $(-2/3,1/3,l)$ at 3.5~T. It is not possible to resolve any clear modulation in the energy, either in the intensity map or after extracting peak positions through taking cuts and performing Gaussian fits to the peak lineshapes (not shown). There is, however, some modulation in the extracted linewidth that could potentially be beyond the experimental error bars, as shown in Fig.~\ref{F:interlayer}(c). Such a linewidth modulation would be expected if the dynamical response contained two unresolved modes with distinct dispersion along $l$. This would be expected in the case of a twinned sample. In particular, the dispersions and dynamical correlations in Eqs.~(\ref{E:lswtinterlayer}) to (\ref{E:Szzinterlayer}) apply for a single crystal with the $P\bar{3}$ structure, but as explained in Appendix~\ref{A:xrays}, a macroscopic size sample is expected to contain co-existing structural twins related by a 2-fold rotation around the $\bm{a}+\bm{b}$ axis, or equivalently the (110) axis. This rotation maps the wave vector ($hkl$) in the axes of the main grain into ($kh\bar{l}$) in the axes of the rotated twin. Therefore, the corresponding dispersions and dynamical correlations for the rotated twin are obtained by replacing $l\rightarrow -l$ in the expression for $\gamma_b$ in Eq.~(\ref{E:gamma_a}). For a twinned sample, this leads to two modes with distinct dispersion along the $l$-direction, one with minimum for $l$ near 1/6 and the other near $-1/6$, illustrated in Fig.~\ref{F:interlayer}(b) by the white and black solid lines, respectively. The two modes cross at $l=0,\pm0.5$, leading to a variation in the effective linewidth of the summed response, as illustrated in Fig.~\ref{F:interlayer}(c); the calculations assumed an equal-weighted average of the two structural twins. 
 
Based on the variation in linewidth, and making the assumption that this is due to two modes dispersing only due to $J_b$, the value $J_b \approx 0.003$~meV is estimated such as to approximately match the overall range of variation of the linewidth in the data and calculation, i.e., about 2.5\% of the value of the dominant nearest-neighbor Ising component $J_z$ refined in Sec.~\ref{S:XXZfit}. The linewidth calculated based on these assumptions is shown by the solid line in Fig.~\ref{F:interlayer}(c) and calculations of the expected inelastic neutron scattering spectrum based on this model are shown in Fig.~\ref{F:interlayer}(b). We note that the model predicts a clear modulation in energy, which is not present in the data. We have verified that adding $J_2=J_b/3$ as used for the magnetic structure model in Fig.~\ref{F:mag_structure}(b) would produce a similar level of agreement with the data. These comparisons suggest that perhaps a more complex model of interlayer interactions would be needed to get closer to the data. There is not, however, sufficient information in the present data to test such a model, so this is considered beyond the scope of this work. This estimated value of $J_b$ is therefore considered as an indicative upper bound on the strength of the interlayer couplings.  

Both because there is no visible modulation in the magnon energies due to interlayer couplings (neither at 3.5~T, nor at the other fields investigated in Horace scans), and because the estimated upper bound of the interlayer couplings is so much smaller than the nearest-neighbor in-plane coupling, their effect is neglected in the rest of this paper. The inelastic neutron scattering data are therefore averaged across the entire range in the $(00l)$ direction to improve counting statistics, and the dynamics are treated as fully two dimensional in all calculations outside this Appendix.

\end{document}